\documentclass[a4letter,10pt,journal]{IEEEtran}

\usepackage{times}
\usepackage{amsmath,amsfonts,amssymb}
\usepackage{tabularx}
\usepackage{slashbox}
\usepackage{color}
\usepackage{fixltx2e}
\usepackage{rotating}

\newcommand{\rank} {\mbox{\rm rank}\,}
\newcommand{\diag} {\mbox{\rm diag}\,}
\newcommand{\nrank} {\mbox{\rm normal rank}\,}

\newtheorem{theorem}{Theorem}[section]
\newtheorem{definition}{Definition}[section]
\newtheorem{proposition}{Proposition}[section]
\newtheorem{lemma}{Lemma}[section]
\newtheorem{corollary}{Corollary}[section]
\newtheorem{example}{Example}[section]
\newtheorem{remark}{Remark}

\begin{document}


\title{\Large \bf On the Zero-freeness of Tall Multirate Linear Systems}

\author{Mohsen Zamani $^{a,}$$^{\ast}$\thanks{$^\ast$Corresponding author. Email: mohsen.zamani@anu.edu.au
\vspace{6pt}}, Giulio Bottegal $^{b}$\vspace{6pt} and  Brian D. O. Anderson $^{a,}$$^{c}$ \\\vspace{6pt}  $^{a}${\em{Research School of Engineering, Australian National University, Canberra, ACT 0200, Australia. (e-mail: \{mohsen.zamani, brian.anderson\}@anu.edu.au)}}\\
$^{b}${\em{ACCESS Linnaeus Centre, School of Electrical Engineering, KTH Royal Institute of Technology, SE-100 44 Stockholm, Sweden. (e-mail: bottegal@kth.se})}\\
$^{c}$ {\em{Canberra Research Laboratory, National ICT Australia Ltd., PO Box 8001, Canberra, ACT 2601, Australia. (e-mail: brian.anderson@anu.edu.au)}} }

\maketitle

\begin{abstract}
In this paper, tall  discrete-time  linear  systems with multirate outputs are studied. In particular, we focus on their zeros. In    systems and control  literature   zeros  of  multirate systems are defined as those  of  their corresponding time-invariant blocked systems. Hence,  the zeros of tall  blocked systems resulting from blocking of linear systems with multirate outputs are mainly explored in this work. We  specifically investigate zeros of tall blocked systems formed  by blocking tall multirate linear  systems with  generic parameter matrices. It is demonstrated that tall blocked systems  generically have no finite nonzero zeros; however, they may have zeros at the origin or at infinity depending on the choice of  blocking delay and the input, state and output dimensions.


\end{abstract}

\section{Introduction}\label{intro}

Multirate linear systems have  been studied in different subdisciplines,  such as sampled-data control \cite{chenB95}, signal processing \cite{vaid93} and econometric modeling \cite{clement2008} for some decades. Especially, with recent theoretical advances   in the field of econometric modeling  (see e.g. \cite{Forni2000}), multirate linear systems analysis  has found more potential applications in `mixed frequency' data analysis; mixed frequency data refers to the fact that in econometric modeling, it is  common  to have some  data which are collected  monthly,  while  other data  may be obtained quarterly or even annually \cite{Schumacher2006}, \cite{clement2008} (in most advanced countries, the number of such time series generally easily exceeds 100). The authors of the present paper have also become interested in  multirate linear systems analysis  while studying generalized dynamic factor models  (GDFMs) \cite{Forni2000}, which are a major tool in the field of econometric modeling. In GDFMs, linear dynamic systems driven by white noise are used to model measured high-dimensional time series, and virtually always, such  systems have a much larger number of outputs than inputs   \cite{Raknerud2007}, \cite{diestler2010} i.e. the systems are tall (if not very tall). Typical research questions include how such models can be identified, and how they can be used for near-term forecasting.

Tall and very  linear multi-rate systems have not been studied in great depth. This paper however does try to formulate some general properties of tall multi-rate systems. Consequently, we do not focus on a particular application problem  but rather on a bigger framework which is the system theoretical issues associated with such systems

As first attempts to understand the properties of tall multirate linear systems, the  authors of  \cite{diestler2010} and  \cite{filler2010} have considered just the single-rate scenario and  shown that  the underlying model is generically \textit{zero-free}. This has the key consequence that identification of the model from measured output data (assuming a white noise input) becomes far simpler than for a normal system, as the system parameters can be identified through linear calculations from the observed data, using a set of equations known as the {\textit{Yule-Walker}} equations \cite{Ltkepohlbook}. A corresponding demonstration till now has been lacking for the  multirate case, and the central task of this paper is to address that shortcoming. Specifically, we show that tall multirate linear systems are generically zero-free, apart possibly for zeros at infinity or zero.

While our prime motivation has been to demonstrate a property which implies, as noted above, substantial simplification in the identification or modelling task, we comment that the result may have separate  importance from a control design perspective;  zeros which are unstable or stable but close to a stability boundary can provide obstructions to the existence of inverses of linear systems and more generally, the design of high performance controllers.    The results of this paper  suggest that, when one is dealing with a generic system, the controller  design may then  be easier if one can add extra sensors to make the system have  more outputs than inputs, and thereby suppress occurrence of any zeros at all, apart possibly from zeros at zero or infinity.

There exists a large number of works in systems and control literature dealing with multirate linear systems; for example,  one can refer to  \cite{bittanti09}, \cite{chenB95}, \cite{Bolzern86}, \cite{Colaneri95}, \cite{Khargoneckar85}, \cite{Colaneri1990}, \cite{Colaneri92} and references listed therein. In order  to deal with this type of system, a technique termed blocking or lifting  has been developed in systems and control \cite{bittanti09}  and signal processing \cite{vaid93} . In systems and control, blocking has been largely used to    transform linear discrete-time periodic systems to linear time-invariant (LTI) systems,  so that analysis and design of the former can be done using   the  well-developed  tools in LTI systems. In particular,  in \cite{bittanti09} and \cite{bittanti86}  the notions of poles and zeros of LTI systems have been extended to linear periodic systems. Moreover, the  authors of  \cite{Grasselli88} and \cite{Bolzern86} have defined  zeros of   multirate linear systems as those  of their corresponding  blocked systems. However, to the  best of our knowledge there are few works on zeros of multirate systems.   Among such works  we should mention  \cite{Bolzern86}, \cite{Grasselli88}, \cite{bittanti09}, \cite{zamani2011}, \cite{chen2010}, \cite{zamani2012acc}, \cite{zamaniscl}, \cite{anderonandzamani2012}.\footnote
{Note that zeros of unblocked LTI systems have been extensively   studied in the literature  (see e.g.  \cite{rosenbrock1974}, \cite{wonhem1979}, \cite{hespanha_book},   \cite{karsanias1979}, \cite{kailath}, \cite{karsanias2002},  \cite{fillerthesis} and  \cite{Christou2010}, \cite{Mitrouli93}).}

References   \cite{Bolzern86} and  \cite{Grasselli88}  have explored  zeros of blocked systems obtained from blocking of linear periodic systems. The results show that the blocked system has a finite zero if it is  obtained from a LTI  unblocked system, and the latter has a finite zero, which is a  form of sufficiency condition.  References  \cite{zamani2011} and \cite{chen2010}  have used different approaches but they have obtained largely  similar results. The results in those references show that a tall  blocked system  has a zero  if and only if its associated LTI unblocked system has a zero. Later, in \cite{zamaniscl} the  authors have obtained more general results by relaxing the assumptions made in   \cite{zamani2011} and \cite{chen2010} on the normal rank and the structure of the transfer function matrices. While references \cite{zamani2011}, \cite{chen2010} and \cite{zamaniscl}  have mainly  considered    LTI  unblocked systems, as opposed to multirate systems, in \cite{zamani2012acc}   zeros of a class of unblocked multirate linear systems  have been explored.   It  has been shown that
the    tall blocked systems obtained from  blocking of multirate systems  with generic parameter matrices have \textit{no finite
nonzero zeros}. Finally, some of the results in \cite{zamani2011} and \cite{zamani2012acc} are reviewed in \cite{anderonandzamani2012}.

The main objective of this paper is to investigate zeros  of  tall blocked systems resulting from blocking of a multirate linear system with a generic choice of parameter matrices appearing in a state-variable description of the system.  The results of this  study  reveal what kind of zeros tall blocked   systems have for almost all choices of parameter matrices. Note that there  are  already some results  in the literature dealing with  zeros of  unblocked tall LTI  systems with generic parameter matrices  \cite{anderson2008}, \cite{filler2010}, \cite{wonhem1979} and \cite{kailath}. However, there has been a gap in the literature  regarding the study of   blocked   systems  formed by blocking of multirate linear system; this process result  in a time-invariant system with relations among the entries of the state-variable matrices of the blocked system, i.e. so that the state-variable matrices are not  fully generic. As mentioned earlier, reference \cite{zamani2012acc} has partially addressed this problem, showing that tall blocked   systems generically have no finite nonzero zeros. However, zeros at the origin and zeros at infinity have not been completely studied in  \cite{zamani2012acc}. Moreover, \cite{zamani2012acc} has mainly focused on situations where the system matrix associated with the blocked system attains full-column normal rank; perhaps surprisingly, this is a significant restriction, and in this paper, there is no such restriction at all.

More precisely, in this paper,  we  provide for the first time a complete analysis of zeros   for  tall  blocked systems obtained by  blocking multirate linear systems with generic parameter matrices.  In particular, as far as zeros at infinity and the origin are concerned, the results of this paper go far beyond the scope of \cite{zamani2012acc}. Here, we show that in general   blocked systems may have zeros at  the origin or  infinity depending on the delay associated with blocking  and the dimension of the input, state and output vectors. Moreover, regarding finite nonzero zeros, the current paper improves some deficiencies in results of  \cite{zamani2012acc} and  shows that tall  blocked systems generically have no finite nonzero zeros.

Since the analysis of zeros  for  tall blocked systems  is quite involved, we consider
three cases separately, that is, 1) finite nonzero system zeros; 2) system zeros at infinity; and 3) system zeros at zero. The next  section of the paper  is focused  on  zeros  of tall blocked systems associated with   finite nonzero zeros. It is explicitly established  that  tall blocked systems  generically have no finite nonzero zeros. As a byproduct in this section, we also establish results on the generic rank of a system matrix resulting from blocking a multi-rate system.  Following this, in  Section \ref{sec:Blocked systems with generic parameters- zeros at origin and infinity}
zeros  of  tall blocked systems are examined at $Z=0$ and $Z=\infty$. It is shown when  tall blocked systems can have a zero at $Z=0$ or $Z=\infty$ and when they are zero-free at those aforementioned points. Finally, Section \ref{sec:conclusion} offers concluding remarks.

\section{Blocked systems with generic parameters- finite nonzero  zeros}\label{sec:Blocked systems with generic parameters-nonzero finite zeros}

In this section, first the formulation of  the problem under study is introduced.   Then  attention is given to the analysis of zeros for  tall  blocked systems  with generic parameters, considering in this section  finite nonzero zeros only.  In the next section,  infinite zeros and zeros at the origin  are explored.

The dynamics  of an  underlying system operating at the highest sample rate  are  defined by
\begin{equation}
\begin{split}\label{unblockedsystem}
x(k+1) &=  A x(k)+B u(k)\\
y(k) &=   C x(k)+Du(k),
\end{split}
\end{equation}
where $x(k)\in \mathbb{R}^{n}$ is the state, $y(k)\in \mathbb{R}^{p}$ the output, and $u(k)\in \mathbb{R}^{m}$ the input.  For this system,  $y(k)$ exists for all $k$, and, separately, can be measured at every time $k$. However, we are also interested in the situation where though  $y(k)$ exists for all $k$, not every entry is measured for all $k$. In particular, we consider the case where   $y(k)$ has components that are observed at different rates. For simplicity, in this paper we  consider a  case where outputs are provided at two rates which we refer to as the fast rate and the slow rate.

Without loss of generality we decompose $y(k)$ as $y(k)=\left[\begin{array}{cc}y^{f}(k)^T & y^{s}(k)^T\end{array}\right]^T$
where the fast part  $y^{f}(k)\in \mathbb{R}^{p_{1}}$ is observed at all $k$,  and  the slow part $y^{s}(k)\in \mathbb{R}^{p_{2}}$ is observed at $k=0,N,2N,\dots$, also   $p_1>0, p_2>0$ and  $p_{1}+p_{2}=p$.
Accordingly, we decompose $C$ and $D$ as

\[
C=\left[\begin{array}{c}C^f\\C^s\end{array}\right],\: D=\left[\begin{array}{c}D^f\\D^s \end{array}\right].
\]

Thus,  the multirate linear system  corresponding to what is measured has the following dynamics:

\begin{equation}
\begin{split}
x(k+1)&=Ax(k)+Bu(k)\:\:\:\:\:\:\:k=0,1,2,\ldots \\
y^f(k)&=C^fx(k)+D^fu(k) \:\: k=0,1,2,\ldots \\
y^s(k)&=C^sx(k)+D^su(k) \:\:\: k=0,N,2N,\ldots
\end{split}\label{eq:multiratesys}
\end{equation}

We have actually $N$ distinct alternative ways to block  the system, depending on how the fast signals are grouped with the slow signals.  Even though these  $N$ different systems share some common poles,  their zeros are not identical in the  whole complex plane (see \cite{bittanti09}, pages 173-179).

We index these systems with an integer  $\tau\in \{1,2,\dots, N\}$, and define
\begin{align}\label{eq:blockedvector}
U_\tau(k) & \triangleq  \left[\begin{array}{c}
u(k+\tau)\\ u(k+\tau+1) \\ \vdots \\ u(k+\tau+N-1)\end{array}\right], \\
Y_\tau(k) & \triangleq  \left[\begin{array}{c}
y^f(k+\tau) \\ y^f(k+\tau+1) \\ \vdots \\ y^f(k+\tau+N-1)\\y^s(k+N)\end{array}\right],\\
x_\tau(k) & \triangleq x(k+\tau),
\end{align}
where $k=0,N,2N,\ldots$.

Then the blocked system $\sum_\tau$ is defined by
\begin{equation}
\begin{split}\label{blockedsystem1}
x_{\tau}(k+N)&=A_{\tau} x_{\tau}(k)+B_{\tau}U_{\tau}(k)\\
Y_{\tau}(k)&=C_{\tau}x_{\tau}(k)+D_{\tau}U_{\tau}(k),
\end{split}
\end{equation}
where
\begin{equation}
\begin{split}
A_{\tau} &\triangleq A^N,\\
B_{\tau} &\triangleq  \left[\begin{array}{ccccc} A^{N-1}B & A^{N-2}B& \ldots& AB & B \end{array}\right],\\
C_{\tau} &\triangleq \left[ \begin{array}{ccccc}\!\!C^{f^T}&\! \! A^TC^{f^T}&\! \!\ldots\! &\! \!A^{(N-1)^{T}}C^{f^T}&\! \!A^{(N-\tau)^{T}}C^{s^T}\end{array}\right]^T\!\!,\\
D_{\tau}&\triangleq \left[\begin{array}{cccc} D^f & 0 & \ldots & 0\\
C^fB& D^f & \ldots & 0 \\
\vdots & \vdots & \ddots & \vdots \\
C^fA^{N-2}B & C^fA^{N-3}B & \ldots & D^f \\ \hline
  & D^s_\tau& \end{array}\right],\\
\end{split}\label{blockedsystemMatrices1}
\end{equation}

where  $ D^s_{\tau}=[C^s A^{N-\tau-1}B\;\ldots\;C^s B\;D^s\;0\;\ldots\;0]$ for $\tau<N$
with $\tau-1$ zero blocks of size $p_{2}\times m$, and when $\tau=N$, it is given by
$D^s_{\tau}=[D^s\;0\;\ldots\;0]$ where there are $N-1$ zero blocks of size $p_{2}\times m$.

Reference \cite{bittanti09} defines   zeros of (\ref{eq:multiratesys}) at time $\tau$ as   zeros of its  corresponding blocked system $\sum_\tau$ \footnote {Zeros of the transfer function defined from (\ref{blockedsystem1}) are identical with those defined here, provided the quadruple $\{A_{\tau},B_{\tau},C_{\tau},D_{\tau}\} $is minimal.}. Hence, in the rest of this section we focus on   zeros  of the blocked system $\sum_\tau$ $\forall \tau \in \{1,2,\ldots,N\}$.

For completeness, we recall the following standard definition \cite{kailath}.

\begin{definition}\label{def:zero}
The finite zeros of the system $\sum_\tau$  are defined to be the finite values of $Z$ for which the rank of the following system matrix falls below its normal rank
\[
M_\tau(Z)=\left[\begin{array}{cc}
ZI-A_\tau & -B_\tau\\
C_\tau & D_\tau\end{array}\right].
\]

Further, $V_\tau(Z)= C_\tau(ZI-A_\tau)^{-1}B_\tau+D_\tau$, $\tau\in \{1,2,\ldots, N\}$, is said to have an infinite zero when $n+\rank(D_{\tau})$, $\tau\in \{1,2,\ldots, N\}$, is less than the normal rank of $M_\tau(Z)$, $\tau\in \{1,2,\ldots, N\}$, or equivalently the rank of $D_{\tau}$, $\tau\in \{1,2,\ldots, N\}$, is less than the normal rank of $V_\tau(Z)$, $\tau\in \{1,2,\ldots, N\}$. \label{them:def11}
\end{definition}
We also provide the following definition for the geometric multiplicity of a zero:
\begin{definition}\label{def:zeromultiplicity}
The geometric multiplicity of a finite zero $Z_0 \in \mathbb C$ is  normal rank of $M_\tau(Z)$- $\rank(M_\tau(Z_0))$. Moreover, the geometric multiplicity of a zero at infinity is  normal rank of $M_\tau(Z)$ $-n-\rank(D_\tau)$.
\end{definition}
In this paper we use the term multiplicity to refer to the geometric multiplicity.

We treat  zeros  of $\sum_\tau$  $\forall \tau \in \{1,2,\ldots,N\}$, under  a genericity  assumption on the matrices of the unblocked system  and a tallness  assumption. Given that $p_1,p_2>0$, it proves convenient to consider a partition of the set of possible values of $p_1$ and $p_2$ defining tallness of the blocked transfer function into two subsets, as follows:

\begin{enumerate}
  \item $p_1>m$.
  \item $ p_1 \le m$, $ Np_1+p_2 >Nm$.
\end{enumerate}

The first case is  common, perhaps even overwhelmingly common,  in econometric modeling but  the second case is  important from a theoretical point of view, and possibly in other applications.
Our results are able to cover both cases, but separate treatment is required. %

\subsection{Case $p_1>m$}

According to Definition \ref{them:def11}, the normal rank for the system matrix of  $\sum_\tau$  $\forall \tau \in \{1,2,\ldots,N\}$, plays an important role in the analysis of its  zeros; thus, we state  the following  straightforward and preliminary  result for the normal rank of $\sum_\tau$  $\forall \tau \in \{1,2,\ldots,N\}$.

\begin{lemma}\label{lem:normalrankp1>m}
For generic choice of the matrices  $\{A,B,C^s,C^f,D^f,D^s\}$, $p_1 \ge m$, the system matrix of $\sum_\tau$  $\forall \tau \in \{1,2,\ldots,N\}$, has normal rank of $n+Nm$.
\end{lemma}

\begin{IEEEproof}
In a generic stetting and with $p_1 \ge m$, the matrix $D^f$ is of full-column rank. So, due to the structure of $D_\tau$  $\forall \tau\in \{1,2,\dots, N\}$, one can easily conclude that $D_\tau$  $ \forall \tau\in \{1,2,\dots, N\}$,  is  of full-column rank as well. Furthermore,
\begin{align}
M_\tau(Z) & =
\left[\begin {array}{cc} ZI-A_\tau & -B_\tau \\
                             C_\tau      & D_\tau \\
                               \end{array}\right]\nonumber\\
          &\!\!\!\!\!\!\!\!\!\!\!\!\!\!\!\! =\!\!  \left[\begin {array}{cc} \!\! I &\!\!\!\! 0 \!\!\\
               \!\!C_\tau(ZI-A_\tau)^{-1}      &\!\!\!\! I\end{array}\!\!\right]\!\!\! \left[\begin {array}{cc}\!\!\! ZI-A_\tau & \!\!\!\!-B_\tau\!\!\!\! \\
              \!\!\! 0      & \!\!\!\!C_\tau(ZI-A_\tau)^{-1}B_\tau\!+\!D_\tau \!\! \!\! \end{array}\right]
\end{align}
Now observe that $M_\tau(Z)$ has $n+Nm$ columns so,  $n+Nm\ge \nrank (M_\tau(Z))=  \nrank(ZI -A_\tau)+ \nrank (C_\tau(ZI-A_\tau)^{-1}B_\tau+D_\tau)\ge n+\rank( \lim_{Z\rightarrow\infty}[C_\tau(ZI-A_\tau)^{-1}B_\tau+D_\tau])=n+ \rank(D_\tau)=n+Nm$. Hence, the normal rank of  $M_\tau(Z)$ equals the number of its columns.
\end{IEEEproof}
In the  situation where  $p_1>m$, obtaining a result on the absence of finite nonzero zeros is now rather trivial, since the blocked system contains a subsystem obtained by deleting some outputs which is provably zero-free.

\begin{theorem}\label{them:finitenonzerozerop1>m}
For a generic choice of  the matrices $\{A,B,C^s,C^f,D^s,D^f\}$, $p_1 > m$, the  system matrix of $\sum _\tau$ $\forall \tau \in \{1,2,\ldots,N\}$,  has  full-column rank for all finite nonzero  $Z$.
\end{theorem}

\begin{IEEEproof}
Define a system matrix $M^f(Z)$ by deleting those rows of $M_\tau(Z)$, $\tau \in \{1,2,\ldots,N\}$,  which contain any entries of $C^s$. Thus $M^f(Z)$ is a system matrix associated with a blocked version of the original system with slow outputs completely discarded, i.e. of a time-invariant and not just periodic system.  With  $p_1>m$, it was shown in  \cite{zamani2011} that  $M^f(Z)$ is generically of  full-column rank for all finite nonzero $Z$. Then it is immediate that $M_\tau(Z)$, $\tau \in \{1,2,\ldots,N\}$,  will be of full-column rank for all  finite nonzero $Z$.
\end{IEEEproof}

\subsection{ Case $p_1 \le m$, $Np_1+p_2>Nm$ }

In the previous subsection the case $p_1>m$ was treated where only considering the fast outputs alone   generically leads to a zero-free blocked system, and the zero-free property is not disturbed by the presence of the further slow outputs.  A different way in which the blocked system will be tall arises when $p_1 \le m$ and  $Np_1+p_2>Nm$.  The main result  of this subsection is to show that
$\sum_\tau$ $\forall \tau \in \{1,2,\ldots,N\}$ with $p_1 \le m$, $Np_1+p_2>Nm$  is again  generically  zero-free. This case is harder to treat; in the conference paper \cite{zamani2012acc}, we treated the case under a restrictive assumption, namely that the system matrix of the blocked system had full column rank, and we shall drop this assumption here. That the system matrix of the blocked system may indeed have less than full column rank, so the extension is warranted,  is exhibited in the following example.
\begin{example}
Consider a tall multi-rate system with $n=1$, $m=3$, $N=2$, $p_{1}=1$,
$p_{2}=5$. Let the parameter matrices for the multirate system
be $A=a$, $B=[b_{1}\,b_{2}\,b_{3}]$, $C^f= c^f,\, C^{s}=[c_{1}^{s}\, c_{2}^{s}\,c_{3}^{s}\, c_{4}^{s}\, c_{5}^{s}]^{T}$, $ D^{f}=[d_{1}^{f}\, d_{2}^{f}\, d_{3}^{f}]$
and
$$D^{s}=\left[\begin{array}{ccc}
d_{11}^{s} & d_{12}^{s} & d_{13}^{s}\\
 \vdots & \vdots & \vdots\\
d_{51}^{s} & d_{52}^{s} & d_{53}^{s}\end{array}\right]\,.$$
All the scalar parameters are generic. We consider $\tau=1$ and  write the associated  system matrix as
\[
M_1(Z)\!\!=\!\!\left[\begin{array}{ccccccc}\!\!
Z-a^{2} & -ab_{1} & -ab_{2} & -ab_{3} & -b_{1} & -b_{2} & -b_{3}\\
c^{f} & d_{1}^{f} & d_{2}^{f} & d_{3}^{f} & 0 & 0 & 0\\
c^{f}a & c^{f}b_{1} & c^{f}b_{2} & c^{f}b_{3} & d_{1}^{f} & d_{2}^{f} & d_{3}^{f}\\
c_{1}^{s}a & c_{1}^{s}b_{1} & c_{1}^{s}b_{2} & c_{1}^{s}b_{3} & d_{11}^{s} & d_{12}^{s} & d_{13}^{s}\\
c_{2}^{s}a & c_{2}^{s}b_{1} & c_{2}^{s}b_{2} & c_{2}^{s}b_{3} & d_{21}^{s} & d_{22}^{s} & d_{23}^{s}\\
c_{3}^{s}a & c_{3}^{s}b_{1} & c_{3}^{s}b_{2} & c_{3}^{s}b_{3} & d_{31}^{s} & d_{32}^{s} & d_{33}^{s}\\
c_{4}^{s}a & c_{4}^{s}b_{1} & c_{4}^{s}b_{2} & c_{4}^{s}b_{3} & d_{41}^{s} & d_{42}^{s} & d_{43}^{s}\\
c_{5}^{s}a & c_{5}^{s}b_{1} & c_{5}^{s}b_{2} & c_{5}^{s}b_{3} & d_{51}^{s} & d_{52}^{s} & d_{53}^{s}\end{array}\!\!\right]\]
It is obvious that first the two rows are (generically) linearly independent. Now consider
rows from 3 to 8; they can be written as a product of matrices $G  \, \bar\Gamma$, with
\[
\begin{split}
G \triangleq \left[\begin{array}{ccccccc}
c^{f} & c^{f} & c^{f} & c^{f} & d_{1}^{f} & d_{2}^{f} & d_{3}^{f}\\
c_{1}^{s} & c_{1}^{s} & c_{1}^{s} & c_{1}^{s} & d_{11}^{s} & d_{12}^{s} & d_{13}^{s}\\
c_{2}^{s} & c_{2}^{s} & c_{2}^{s} & c_{2}^{s} & d_{21}^{s} & d_{22}^{s} & d_{23}^{s}\\
c_{3}^{s} & c_{3}^{s} & c_{3}^{s} & c_{3}^{s} & d_{31}^{s} & d_{32}^{s} & d_{33}^{s}\\
c_{4}^{s} & c_{4}^{s} & c_{4}^{s} & c_{4}^{s} & d_{41}^{s} & d_{42}^{s} & d_{43}^{s}\\
c_{5}^{s} & c_{5}^{s} & c_{5}^{s} & c_{5}^{s} & d_{51}^{s} & d_{52}^{s} & d_{53}^{s}\end{array}\right]
\end{split}
\]
and $\bar \Gamma \triangleq \diag(a,b_{1},b_{2},b_{3},I_{3})$.
The matrix $G$ has rank at most 4; hence, with generic parameter matrices the normal rank of $M_1(Z)$ equals 6   and thus $M_1(Z)$ cannot attain full-column normal rank.
\end{example}

In the next part of this subsection, we first characterise the normal rank of $M_{\tau}(z)$; following that, we turn to the question of zero existence.

\begin{proposition} \label{prop:rankDtau}
Consider the system $\sum_\tau$ $\forall \tau \in\{1,2,\ldots,N\}$, with
$p_1 \leq m$, $Np_1+p_2>Nm$ and generic values of the defining matrices
$\{A,\,B,\,C^f,\,C^s,\,D^f,\,D^s\}$. Then
\begin{enumerate}
\item if $n\leq (N-\tau)(m-p_1)$, the matrix $D_\tau$ has rank equal to $(N-1)p_1+m+n$;
\item if $n > (N-\tau)(m-p_1)$, the matrix $D_\tau$  has rank equal to $(\tau-1)p_1+(N-\tau+1)m$.
\end{enumerate}
\end{proposition}
\begin{IEEEproof}
Refer to the appendix for a proof.
\end{IEEEproof}

Now the general result on the normal rank of $M_{\tau}(z)$ is as follows:

\begin{theorem} \label{less_than_normal_rank}
Consider the system $\sum_\tau$, $\tau \in \{1,2,\ldots,N\}$, with
$p_1 < m$, $Np_1+p_2>Nm$ and generic values of the defining matrices
$\{A,\,B,\,C^f,\,C^s,\,D^f,\,D^s\}$.  Then the normal rank of the system matrix $M_\tau(Z)$ is equal to:
\begin{enumerate}
\item $(N-1)p_1+m+2n$, if  $n < (N-1)(m-p_1)$;
\item $n+Nm$, if  $n \geq (N-1)(m-p_1)$.
\end{enumerate}
\end{theorem}

\begin{IEEEproof}
The proof is provided in the appendix.
\end{IEEEproof}
We return to the main task of studying the zeros of the blocked system. For this purpose, we first review briefly properties of the Kronecker canonical form of a matrix pencil. The system matrix of  $\sum _\tau$ $\forall \tau \in \{1,2,\ldots,N\}$ is actually a matrix pencil, and  the  Kronecker canonical form turns out to be a very  useful tool  to obtain insight into the zeros  of    (\ref{blockedsystem1})  and the structure of the kernels associated with those zeros.

The main theorem on   the Kronecker canonical form of a  matrix pencil is obtained  from \cite{vandooren1979}.

\begin{theorem}\cite{vandooren1979}\label{them:vandoren}
Consider a matrix pencil $zR+S$. Then under the equivalence defined using pre- and postmultiplication by nonsingular constant matrices $\widetilde{P}$ and $\widetilde{Q}$, there is a
canonical quasidiagonal form:
\begin{equation}
\widetilde{P}(zR+S)\widetilde{Q}\!=\!\diag \![L_{\epsilon_1},\dots,L_{\epsilon_r}, \tilde L_{\eta_1},\dots,\tilde L_{\eta_s},zN-I, zI-K]
\end{equation}
where:
\begin{enumerate}
\item
$L_{\mu}$ is the $\mu \times (\mu+1)$ bidiagonal pencil
\begin{equation}
\left[\begin{array}{cccccc}
z&-1&0&\dots&0&0\\
0&z&-1&\dots&0&0\\
\vdots&\vdots&&&&\vdots\\
0&0&0&\dots&z&-1
\end{array}\right].
\end{equation}
\item
$\tilde L_{\mu}$ is the $(\mu+1)\times \mu$ transposed bidiagonal pencil
\begin{equation}
\left[\begin{array}{ccccc}
-1&0&\dots&0&0\\
z&-1&\dots&0&0\\
\vdots&&&&\vdots\\
0&0&\dots&z&-1\\
0&0&\dots&0&z
\end{array}\right].
\end{equation}
\item
$N$ is a nilpotent Jordan matrix.
\item
$K$ is in Jordan canonical form.
\end{enumerate}

\end{theorem}
Furthermore, the possibility that  $\mu=0$ exists. The associated $L_0$ is deemed to have  a column but not a row and $\tilde{L}_0$ is deemed  to have a row but not a column, see \cite{vandooren1979}.

The following corollary can be directly derived easily from the  above  theorem and provides detail about the  vectors in the null space of  the Kronecker canonical form. Because the matrices $\tilde P$ and $\tilde Q$ are nonsingular, it is trivial to translate these properties back to an arbitrary matrix pencil, including a system matrix.

\begin{corollary}
With the same hypothesis as Theorem \ref{them:vandoren}, and with $\Lambda(K)$ denoting the set of eigenvalues of $K$, the following hold:
\begin{enumerate}
\item
For all $z \notin \Lambda(K)$, the kernel of the Kronecker canonical form has  dimension equal to the number of matrices $L_{\mu}$ appearing in the form; likewise  the co-kernel dimension is determined by the number of matrices $\tilde L_{\mu}$.
\item
The vector $[1\;z\;z^2\dots z^{\mu}]^T$ is the generator of the kernel of $L_{\mu}$,  a set of vectors  $$\left[0 \dots 0 \; 1\;z\;z^2\dots z^{\mu}\;0 \dots 0 \right]^T$$  are generators for the kernel of the whole canonical form which depend continuously on $z$, provided that $z \notin \Lambda(K)$; when $z \in \Lambda(K)$, the vectors form a subset of a set of generators.
\item
When $z \in \Lambda(K)$ equals an eigenvalue of $K$, the dimension of the kernel jumps by the geometric multiplicity of that eigenvalue, the rank of the pencil drops below the normal rank by that geometric multiplicity, and there is an additional vector or vectors in the kernel apart from those defined in point 2,  which are of the form $[0\;0\dots v^T]^T$, where $v$ is an eigenvector of $K$. Such a vector is orthogonal to all vectors in the kernel which are a linear combination of the generators listed in the previous point.
\item
Let $\lambda_0\in\Lambda(K)$  the associated kernel of the matrix pencil can be generated by two types of vectors: those which are the limit of the generators defined by adding extra zeros to vectors such as  $[1\;\lambda_0\;\lambda_0^2\dots,\lambda_0^{\mu}]^T$ (these being the limits of the generators when $z\neq \lambda_0$ but continuously approaches $\lambda_0$),  and those obtained by adjoining zeros to the eigenvector(s)  corresponding to $\lambda_0$,  the latter set being orthogonal to the former set.
\end{enumerate}
\end{corollary}

In the  rest of this subsection, we explore     zeros of $M_\tau(Z)$  $\forall \tau \in \{1,2,\ldots,N\}$. To achieve this,
we first focus on the particular case of $M_1(Z)$. Later,  we introduce  the main result for  zeros of   $M_\tau(Z)$ $\forall \tau \in \{1,2,\ldots,N\}$.

We begin by studying a square matrix generated from certain rows of $M_1(Z)$; these are the rows remaining after    excluding certain output variables from consideration.   To this end, we argue first that the first $n+Np_1$ rows of $M_1(Z)$ are linearly independent.
For the submatrix formed by these rows is the system matrix of the blocked system obtained by blocking the fast system defined by $\{A,B,C^f,D^f\}$,
and accordingly has full-row normal rank, since the unblocked system is generic and square or fat under the condition $p_1\leq m$. Now define the square submatrix of $M_1(Z)$:

\begin{equation}\label{N(Z)}
N(Z)\triangleq\left[\begin{array}{cc}
ZI- A_1&-B_1\\
\mathcal C_1&\mathcal D_1
\end{array}\right],
\end{equation}
such that $\text{normal rank} (N(Z))= \text{normal rank}(M_1(Z))$, by including the first $n+Np_1$ rows of $M_1(Z)$ and followed by appropriate other rows of $M_1(Z)$ to meet the normal rank and squareness requirements.
Note that there exists a permutation matrix $P$ such that
\begin{equation}\label{M(Z)partioned}
P M_1(Z) = \left[\begin{array}{c} N(Z) \vspace{1mm} \\  \mathcal C_2 \:\:\:\: \mathcal D_2 \end{array} \right], \,
\end{equation}
where $\mathcal C_2$ and  $\mathcal D_2$ capture those  rows of $C_1$ and $D_1$ that are  not included in  $\mathcal C_1$ and  $\mathcal D_1$, respectively.

The zero properties  of $N(Z)$ are  studied in the following proposition (we will build on them to obtain the zero properties of $M_1(Z)$).

\begin{proposition}
Let the matrix $N(Z)$ be the  submatrix of $M_1(Z)$ formed via the procedure described.  Then for  generic values of the matrices $A, B$, etc. with  $p_1 \le m$ and   $Np_1+p_2>Nm$,  for any finite $Z_0$ for which the matrix $N(Z_0)$ has less rank than its normal rank, its rank is one less than its normal rank.
\end{proposition}

\begin{IEEEproof}
We distinguish two cases, $p_1=m,\; p_1<m$. In case $p_1=m$, then $N(Z)$ is the system matrix for the system obtained by blocking the original system with  slow outputs discarded. As such, the blocked system zeros are precisely the $N$-th powers of the unblocked system zeros \cite{zamani2011}. For generic coefficient matrices,  the unblocked system will have $n$ distinct zeros; then the blocked system will have the same property. Further, the unblocked system will generically have a nonsingular direct feedthrough matrix, as will  then the blocked system, so that $\mathcal D_1$ can be assumed to be nonsingular.  It follows then that the zeros of the system with system matrix $N(Z)$ are identical with the eigenvalues of $ A_1- B_1\mathcal D_1^{-1}\mathcal C_1$,  which are then distinct,  and since this matrix is $n \times n$, the eigenvector associated with each zero will be uniquely defined to within a scaling constant. It follows easily that there is a unique vector (to within scaling) in the kernel of $N(Z_0)$ where $Z_0$ is the zero of the blocked system.

We turn therefore to the case $p_1<m$. We study the co-kernel of  $N(Z_0)$. Let $Z_1,Z_2,\dots,$ be a sequence of complex numbers such that (a) $Z_i\rightarrow Z_0$ and (b) $\rank (N(Z_i))$ equals the normal rank of $N(Z)$. From what has been described earlier using the Kronecker canonical form, we know that the sequence of co-kernels of $N(Z_i)$ converges, say to $\mathcal K$, with any vector in this limit also in the co-kernel of $N(Z_0)$. In addition, since $N(Z_0)$ has lower rank than the  normal rank of $N(Z)$, the co-kernel, call it $\mathcal {\bar K}$, will be strictly greater than $\mathcal K$. Suppose its dimension is at least two more than that of $\mathcal K$. We shall show this situation is nongeneric.

Select two vectors $w_1,w_2$ which are in $\mathcal {\bar K}$ and which are orthogonal to $\mathcal K$. Then it is evident that there are two vectors call them $v_1,v_2$, constructed from linear combinations of $w_1,w_2$, which belong to $\mathcal {\bar K}$, which are still orthogonal to  $\mathcal K$, and which for some pair $r<s$ have 1 and 0 in the $r$-th entry and 0 and 1 in the $s$-th entry respectively. Choose $v_1,v_2$ so that firstly, $s$ is maximal, and secondly, for that $s$ then $r$ is  maximal. It is not difficult to see that this means that $v_1$ has zero entries beyond the $r$-th and $v_2$ has zero entries beyond the $s$-th.

Now again we must consider two cases. Suppose firstly that $s$ obeys $n+Np_1+1\leq s\leq n+Nm$; in forming the product $v_2^TN(Z_0)$, the $s$-th entry of $v_2$  will be multiplying entries of $N(Z_0)$ defined using $C^s, A, B, D^s$.  Consider an entry in the $s$-th row of $N(Z_0)$ and in the last $m$ columns.  Such an entry is an entry of $D^s$, and is independent of all other entries in $N(Z_0)$. Suppose this entry of $D^s$ is continuously perturbed by a small amount. Then clearly $v_1$  remains in the co-kernel of $N(Z_0)$ but $v_2$ cannot.

The particular values of $Z$ for which $N(Z)$ has rank less than its normal rank, i.e. the zeros of $N(Z)$, will depend continuously on the perturbation.

Accordingly, with a small enough perturbation, those not equal before perturbation to $Z_0$ will never change to $Z_0$, and it is therefore guaranteed that with a small enough nonzero perturbation,  the co-kernel of $N(Z_0)$ is reduced by one in dimension, though never to zero. If the original (before perturbation) co-kernel $\mathcal {\bar K}$ had dimension greater than two in excess of the dimension of $\mathcal K$ , and the excess after perturbation is still greater than one, the argument can be repeated. Eventually, the co-kernel of $N(Z_0)$ will have an excess dimension over $\mathcal K$ of 1, i.e. $N(Z_0)$ will have rank one less than the normal rank of $N(Z)$.

Now suppose that $s$ obeys $s\leq n+Np_1$. Then the last $N(m-p_1)$ entries of each of $v_1,v_2$ are zero. Remove these entries to define  two linearly independent vectors $\tilde v_1, \tilde v_2$ of length $n+Np_1$,  which evidently satisfy

\begin{equation}\label{eq:fat}
\tilde v_i^T
\left[\begin{array}{cccc}
ZI_n-A^N&-A^{N-1}B&\dots&-B\\
C^f&D^f&&0\\
\vdots&\vdots&\ddots&\vdots\\
C^fA^{N-1}&C^fA^{N-2}B&\dots&D^f
\end{array}\right]
=0,  \:\:i=1,2.
\end{equation}

The above equation contains  a fat system matrix, corresponding to a blocked version of a fat time-invariant unblocked system. It can be concluded easily form the results provided in \cite{zamani2011} that  for generic values of the underlying matrices, there can be no $Z_0$ for which an equation such as (\ref{eq:fat}) can even hold for a single nonzero $\tilde v_i$, let alone two linearly independent ones. This ends the proof.
\end{IEEEproof}

The result of the  previous proposition, although restricted to $\tau =1$, enables us to establish the main  result of this section,   applicable for any $\tau$.  Before we state the main theorem we need to recall the following lemma from   \cite{chen2010} and \cite{Colaneri95}.

\begin{lemma}\label{lem:reachability}
The pair $(A,B)$ is reachable if and only if the  pair $(A_\tau,B_\tau)$  $\forall \tau \in \{1,2,\ldots,N\}$ is reachable.
\end{lemma}

\begin{theorem}
Consider   the system $\sum_\tau$, $\forall \tau \in \{1,2,\ldots,N\}$, with  $p_1 \le m$,  and  $Np_1+p_2>Nm$. Then  for generic values of the defining matrices $\{A, B, C^f, D^f, C^s, D^s\}$ the system matrix $M_\tau(Z)$   $\forall \tau \in\{1,2,\ldots,N\}$,  has  rank equal to its normal rank  for all finite nonzero values of  $Z_0$, and accordingly $\sum_\tau$ has no finite nonzero zero. \label{them:infinitezerop1<mbutnp1+p2>nm}
\end{theorem}

\begin{IEEEproof}
We first focus on the case $\tau=1$. Now,  apart from
the $p_2-N(m-p_1)$ rows of the $C^s,D^s$  which do not enter the
matrix $N(Z)$ defined by (\ref{N(Z)}),  choose generic values for
the defining matrices, so that the conclusions of the preceding
proposition are valid.

Let $Z_{a},Z_{b},\dots $ be the finite set of $Z$ for which $N(Z)$ has less rank than its normal rank (the set may have less than $n$ elements, but never has more),
and let $w_{a},w_b,\dots$ be vectors which are  in the corresponding kernels ({\it{not co-kernels}}) and orthogonal to the subspace in the kernel obtained from the
limit of the kernel of $N(Z)$ as $Z\rightarrow Z_a,Z_b,\dots$ etc.  Now, due
to the facts  that $M_1(Z)$ and $N(Z)$ have the same normal rank and     relation (\ref{M(Z)partioned}) holds, it follows that for generic $Z$, the kernels of $M_1(Z)$ and $N(Z)$ are identical (and may be both empty). Hence  one can conclude that the subspace in the kernel obtained from the limit
of the kernel of $N(Z)$ as $Z$  approaches any of $Z_a,Z_b,\dots$ etc. coincides  with the subspace in the kernel obtained from the limit of
the kernel of $M_1(Z)$ as $Z \rightarrow $ zeros of $M_1(Z)$.

Now, to obtain a contradiction, we  suppose that the system matrix
$M_1(Z)$ is such that, for  $Z_0\neq 0$, $M_1(Z_0)$ has rank less
than its normal rank, i.e. the dimension of its kernel increases. Since the kernel of $M_1(Z_0)$ is a subspace of the kernel
of $N(Z_0)$, $Z_0$ must coincide with  one of the values of
$Z_{a},Z_{b},\dots $ and  the rank of $M_1(Z_0)$ must be only one
less than its normal rank; moreover, there must exist an associated
nonzero $w_1$ unique up to a scalar multiplier,  in the kernel of $M_1(Z_0)$ which is orthogonal to the
limit of the kernel of $M_1(Z)$ as $Z\rightarrow Z_0$. Then $w_1$ is
necessarily in the kernel of $N(Z_0)$, orthogonal to the limit of the kernel of $N(Z)$  as
$Z\rightarrow Z_0$  and thus $w_1$ in fact
must coincide to within a nonzero multiplier with one of the vectors
$w_a,w_b,\dots$.

Write this $w_1$ as $w_1=\left[\begin{array}{ccccc}
x_1^T & u_1^T & u_2^T & \ldots & u_{N}^T \end{array} \right]^T$ and suppose the input sequence $u(i)=u_i$ is applied for $i=1,2\dots,N$ to the original system, starting in initial state $x_1$ at time $1$. Let $y^f(1), y^f(2), \dots$ denote the corresponding fast outputs and $y^s(N)$ the slow output at time $N$. Break this up into two subvectors, $y^{s1}(N), y^{s2}(N)$, where $y^{s1}(N)$ is associated with those rows of $C^s$, $D^s$  which are included in  $\mathcal C_1$, $\mathcal D_1 $ (see \eqref{N(Z)}) and $y^{s2}(N)$ is related with the remaining rows of $C^s$ and  $D^s$ . We have $N(Z_0)w_1=$
\begin{align}
&\left[\!\!\begin{array}{ccccc}
Z_0I_n-A^N&-A^{N-1}B&-A^{N-2}B&\dots&-B\\
C^f&D^f&0&\dots&0\\
C^fA&C^fB&D^f&\dots&0\\
\vdots&\vdots&\vdots& &\vdots\\
C^fA^{N-1}&C^fA^{N-2}B&C^fA^{N-3}B&\ldots&D^f\\
C^{s1}A^{N-1}&C^{s1}A^{N-2}B&C^{s1}A^{N-3}B&\ldots&D^{s1}
\end{array}\!\!\right]w_1 \nonumber\\
&=\left[\begin{array}{c}
Z_0x_1-x(N+1)\\y^f(1)\\y^f(2)\\\vdots\\ y^f(N)\\y^{s1}(N)\end{array}\right]=0.
\end{align}

Now it must be true that $x_1\neq0$. For otherwise, we would have $N(Z)w_1=0$ for all $Z$, which would violate assumptions. Since also $Z_0\neq 0$, there must hold $x(N+1)\neq 0$. Hence there cannot hold both $x(N)=0$ and $u(N)=0$. Consequently, we can always find $C^{s2},D^{s2}$ such that $y^{s2}(N)=C^{s2}x(N)+D^{s2}u(N)\neq 0$, i.e. the slow output value is necessarily nonzero, no matter whether $w_1=w_a,w_b,$ etc. Equivalently, the equation $[\mathcal C_2\;\;\mathcal D_2]w_1=0$ cannot hold. Hence,  if $M_1(Z)$ defines a system with a finite zero and it is nonzero,  this is a nongeneric situation.   Hence, $M_1(Z)$ generically has rank equal to its normal rank  for all finite nonzero $Z$.   It now remains to show that this property carries over to all  $M_\tau(Z)$, $\tau \in  \{2,3,\ldots,N\}$.  First, note that the  pair $(A,B)$ is generically reachable; then by Lemma \ref{lem:reachability} the pair $(A_\tau,B_\tau)$,  $ \forall \tau \in \{1,2,\ldots,N\}$, is also reachable.  Consider $Z_\zeta \in \mathbb{C}-\{0,\infty\}$; if $Z_\zeta$ does not coincide   with any eigenvalue  of $A_\tau$ then

\begin{equation}\label{eq:rank(z_zeta)}
\rank(M_\tau(Z_\zeta))=n+\rank(V_\tau(Z_\zeta)).
\end{equation}
Hence, using the result of Proposition \ref{th:constant_normal_rank} (see the appendix),  it is immediate that $\rank(M_\tau(Z_\zeta))=\rank(M_{\tau+1}(Z_\zeta))$. If $Z_\zeta$ does  coincide with  an   eigenvalue  of $A_\tau$ then $\rank(V_\tau(Z_\zeta))$ is ill-defined. However, since  zeros of $M_\tau(Z)$,  $\tau\in\{1,2\ldots,N\}$,   are invariant under  state feedback and the  pair $(A_\tau,B_\tau)$ is reachable,   one can  easily find a state feedback to shift  that eigenvalue   \cite{Zhou1996} and then (\ref{eq:rank(z_zeta)}) is a well-defined equation and $\rank(M_\tau(Z_\zeta))=\rank(M_{\tau+1}(Z_\zeta))$. Thus, we can conclude that all  $M_\tau(Z)$, $\tau \in \{1,2,\ldots,N\}$ generically have no finite nonzero zeros. This ends the proof.
\end{IEEEproof}

\section{Blocked systems with generic parameters- zeros at the origin and infinity}\label{sec:Blocked systems with generic parameters- zeros at origin and infinity}

In the previous  section zeros   of tall  blocked systems with generic parameters for the  choice of finite nonzero zeros were studied. In this  section zeros  of the latter systems are investigated for choices of zeros at zero and infinity. As in the previous section, it is convenient to break up our examination of tall systems into  separate cases based on the relation between $p_1$ and $m$.

We first state the following result which, perhaps surprisingly, relates zeros of the system $\sum_\tau$ at infinity to zeros of the system $\sum_{N-\tau+1}$ at the origin and conversely.
\begin{lemma} \label{th:zeros_zero_infinity}
Consider the family of systems $\sum_\tau$  $\forall \tau \in \{1,2,\ldots,N\}$, where the defining matrices $\{A, B, C^f, D^f, C^s, D^s\}$ assume  generic values. Then the following fact holds:
$\sum_\tau$ has $\kappa$ zeros at $Z=0$ and $\mu$ zeros at $Z=\infty$ if and only if $\sum_{N-\tau+1}$ has  $\mu$ zeros at $Z=0$ and $\kappa$ zeros at $Z=\infty$.
\end{lemma}

\begin{IEEEproof}
Consider a reverse-time  description of the system
\eqref{eq:multiratesys}, namely

\begin{equation}
\begin{array}{rl}
x(k-1)=& \!\!\!A^{-1}x(k)-A^{-1}Bu(k-1) \,\,,\,  k=1,2,\ldots\\
y^f(k-1)=&\!\!\!C^fx(k-1)+D^fu(k-1) \,\,,\, k=1,2,\ldots  \\
 = &\!\!\!C^fA^{-1}x(k) + (D^f - C^fA^{-1}B)u(k-1)     \\
y^s(k-1)=&\!\!\!C^sx(k-1)+D^su(k-1) \,\,,\, k=1,N\!+\!1,\ldots   \\
 = &\!\!\!C^sA^{-1}x(k) + (D^s - C^sA^{-1}B)u(k-1)
\end{array}\label{eq:multiratesysback}
\end{equation}

and define the following matrices
\begin{equation}
\begin{array}{ll}
\tilde A \triangleq A^{-1} & \quad \tilde B \triangleq -A^{-1}B \\
\tilde C^f \triangleq C^fA^{-1} & \quad \tilde D^f \triangleq D^f -
C^fA^{-1}B \\
\tilde C^s \triangleq C^sA^{-1} & \quad \tilde D^s \triangleq D^s -
C^sA^{-1}B \,
\end{array}
\end{equation}
which are still in a generic setting since the genericity
of $\{A, B, C^f, D^f, C^s, D^s\}$ is assumed.  Note that the matrix $A^{-1}$ is well-defined, since $A$ is generically full rank.   Recall  the
blocking procedure introduced in \eqref{eq:blockedvector} for a given value of $\tau$;  we can  obtain  the blocked time-invariant system associated with the system (\ref{eq:multiratesysback}) (again a reverse-time system) as
\begin{equation}
\begin{split} \label{blockedsystem2}
x_{\tau}(k-N)&=\tilde A_{\tau} x_{\tau}(k)+ \tilde B_{\tau}U_{\tau}(k-N)\\
Y_{\tau}(k-N)&= \tilde C_{\tau}x_{\tau}(k)+ \tilde
D_{\tau}U_{\tau}(k-N),
\end{split}
\end{equation}

where $k=N,2N,\ldots$, and
\begin{equation}
\begin{split}
\tilde A_{\tau} &\triangleq \tilde A^N,\\
\tilde B_{\tau} &\triangleq  \left[\begin{array}{ccccc}  \tilde B & \tilde A \tilde B & \ldots&  \tilde A^{N-2} \tilde B&   \tilde A^{N-1} \tilde B\end{array}\right],\\
\tilde C_{\tau} &\triangleq \left[ \begin{array}{cccc} \tilde A^{(N-1)^{T}} \tilde C^{f^T} & \ldots  & \tilde C^{f^T}& \tilde A^{(\tau-1)^{T}} \tilde C^{s^T} \end{array}\right]^T,\\
\tilde D_{\tau}&\triangleq \left[\begin{array}{cccc}\tilde D^f & \ldots & \tilde C^f \tilde A^{N-3} \tilde B& \tilde C^f \tilde A^{N-2} \tilde B   \\
\vdots & \ddots & \vdots   & \vdots \\
0 & \ldots & \tilde D^f & \tilde C^f \tilde B \\
0 & \ldots & 0& \tilde D^f \\
\hline & &\tilde{D}_\tau^s   \end{array}\right].\\
\end{split}\label{blockedsystemMatrices2}
\end{equation}
In the latter expression, when $\tau > 1$ the matrix $\tilde{D}_\tau^s$ is equal to $\begin{bmatrix} 0 & \ldots & 0 & \tilde D^s &  \ldots & \tilde C^s \tilde A^{\tau-2} \tilde B \end{bmatrix}$, with $N-\tau$ zero blocks of size $p_2 \times m$, while, when $\tau = 1$, it becomes $\begin{bmatrix} 0 & \ldots & 0 & \tilde D^s \end{bmatrix}$. Now let us introduce the $N$-step backward operator $\zeta$, such that $\zeta x(k) = x(k-N)$. Then  the transfer function  $\tilde V_\tau(\zeta) \triangleq \tilde C_{\tau}(\zeta I - \tilde A_{\tau})^{-1} \tilde B_{\tau} +
\tilde D_{\tau}$ is  associated with the  blocked system (\ref{blockedsystem2}). It can be easily checked through simple computations that this transfer function is
connected to the   transfer function $V_\tau(Z)$ associated with the system $\sum_\tau$  at the points zero and infinity through the equalities
\begin{equation} \label{eq:prop_zero_infinity}
\tilde V_\tau (0) = \lim_{Z \rightarrow \infty} V_\tau(Z) \qquad \lim_{\zeta \rightarrow \infty}\tilde V_\tau(\zeta) =  V_\tau(0).
\end{equation}
Define the system matrix associated with the system \eqref{blockedsystem2} as
\begin{equation}
\tilde M_\tau(\zeta) \triangleq \left[\begin{array}{cc}
\zeta I- \tilde A_\tau & - \tilde B_\tau\\
\tilde C_\tau & \tilde D_\tau\end{array}\right].
\end{equation}
For our purpose in this paper,  we define the  following equalities
\begin{equation}
\begin{split}
\label{eq:prop_zero_infinity2a}
\rank(\lim_{Z \rightarrow \infty} M_\tau(Z)) &\triangleq n+\rank (D_\tau) \\
\rank(\lim_{\zeta \rightarrow \infty}\tilde M_\tau(\zeta)) &\triangleq  n+\rank (\tilde D_\tau)
\end{split}
\end{equation}

Then  using the equation \eqref{eq:prop_zero_infinity} one can write
\begin{equation}
\begin{split}\label{eq:prop_zero_infinity2}
\rank(\lim_{Z \rightarrow \infty} M_\tau(Z)) &= \rank (\tilde M_\tau(0))\\
\rank(\lim_{\zeta \rightarrow \infty}\tilde M_\tau(\zeta)) &=  \rank(M_\tau(0)).
\end{split}
\end{equation}

Again, note that the above equalities are well-defined since, due the genericity assumption of the  matrix $A$, the  matrices  $A_\tau$ and $\tilde A_\tau$ do not have any eigenvalues at the origin. Now, by comparing  \eqref{blockedsystemMatrices1} and \eqref{blockedsystemMatrices2},  one can verify that there exist permutation matrices  $Q_1$ and $Q_2$ such that   $Q_1M_\tau(\zeta)Q_2=\Psi_\tau(\zeta)$, and $\Psi_\tau(\zeta)$ is exactly  $M_{N-\tau+1}(Z)$ when $\tilde A, \tilde B,\tilde C^f,\tilde C^s,\tilde D^f, \tilde D^s, \zeta$ are  replaced by $A,B, C^f,D^f,C^s,D^s, Z$, accordingly. Since the parameter matrices $A,B,C,D$ assume generic values, we have the following equalities
\begin{equation}
\begin{split} \label{eq:zetavsz2}
\rank(\lim_{\zeta \rightarrow \infty} \tilde M_\tau(\zeta)) &= \rank(\lim_{Z \rightarrow \infty} M_{N-\tau+1}(Z))\\
\rank(\tilde M_\tau(0))&=  \rank (M_{N-\tau+1}(0))
\end{split}
\end{equation}

Then, by combining equations (\ref{eq:prop_zero_infinity2}) and (\ref{eq:zetavsz2}) we obtain
\begin{equation} \label{eq:z=inf}
\rank (\lim_{Z \rightarrow \infty}   M_\tau(Z)) = \rank(M_{N-\tau+1}(0))
\end{equation}
and
\begin{equation}\label{eq:z=0}
  \rank (M_\tau(0)) = \rank( \lim_{Z \rightarrow \infty}   M_{N-\tau+1}(Z)).
\end{equation}
Thus,  by using equations (\ref{eq:z=inf}), (\ref{eq:z=0}) and the fact that the normal rank of $M_\tau(Z)$ does not depend on $\tau$ (see Proposition \ref{th:constant_normal_rank} in Appendix), the conclusion of the lemma readily follows.
\end{IEEEproof}

\subsection{Case $p_1>m$}
\begin{theorem}
For a generic choice of  the matrices $\{A,B,C^s,C^f,D^s,D^f\}$, $p_1 > m$, the system matrix of  $\sum_\tau$  $\forall \tau \in \{1,2,\ldots,N\}$, has  full-column rank at  $Z=0$ and $Z=\infty$, and accordingly $\sum_\tau$ has no zero at $Z=0$ and $Z=\infty$.
\end{theorem}

\begin{IEEEproof}
We first consider the zeros at $Z=0$. It was shown in \cite{zamani2011} that  $M^f(0)$,  where the  system matrix $M^f(0)$  can be formed by deleting rows of $M_1(0)$ which are  related to $C^s$ and $D^s$, has  full-column rank at  $Z=0$ for generic parameter matrices $A,B$, etc. Then it is immediate that $M_\tau(0)$  $\forall \tau \in \{1,2,\ldots,N\}$ has full-column rank, implying that the system $\sum_\tau$  has no zero at $Z=0$.
Next, consider zeros at infinity. Using Lemma \ref{th:zeros_zero_infinity}, it follows that $M_\tau(Z)$ $\forall \tau \in \{1,\,\ldots,\,N\}$ is full-column rank. Hence, $\sum_\tau$ has no zeros at infinity.
\end{IEEEproof}

\subsection{Case $p_1 \le  m$, $Np_1+p_2>Nm$}

%
%

As in the previous subsection, we study zeros  of tall blocked systems at infinity and the origin. We shall start with the former. According to Definition \ref{def:zero}, the rank of matrix $D_\tau$ plays a crucial role in the determination of the zeros at infinity. We now use the result of Proposition \ref{prop:rankDtau} to determine the multiplicity of zeros at infinity.

\begin{theorem}\label{thm:zeroatinfiityp1<m}
Consider   the system $\sum_\tau$  $\forall \tau \in \{1,2,\ldots,N\}$, with  $p_1 \leq m$ and  $Np_1+p_2>Nm$.  Assume that the defining matrices $\{A, B, C^f, D^f, C^s, D^s\}$ take generic values. Then $M_\tau(Z)$    has    zeros at $Z=\infty$ with multiplicity equal to:
\begin{enumerate}
\item $0$ if $n \leq (N-\tau)(m-p_1)$;
\item $n-(N-\tau)(m-p_1)$ if $(N-\tau)(m-p_1) < n \leq (N-1)(m-p_1)$;
\item $(\tau-1)(m-p_1)$ if $n > (N-1)(m-p_1)$.
\end{enumerate}
\end{theorem}
\begin{IEEEproof}
Denote by $\sigma$ the multiplicity of zeros at infinity. Then, by Definition \ref{def:zeromultiplicity} we have $\sigma = \nrank M_\tau(Z) - n - \rank D_\tau$. Consider the following cases.
\begin{enumerate}
\item $n \leq (N-\tau)(m-p_1)$. From Theorem \ref{less_than_normal_rank} we have that $\nrank M_\tau(Z) = (N-1)p_1+m+2n$, while Proposition \ref{prop:rankDtau} yields that $\rank D_\tau = (N-1)p_1+m+n$. Then we easily conclude that $\sigma=0$.
\item $(N-\tau)(m-p_1) < n \leq (N-1)(m-p_1)$. From Theorem \ref{less_than_normal_rank} we still have that $\nrank M_\tau(Z) = (N-1)p_1+m+2n$, while now Proposition \ref{prop:rankDtau} yields $\rank D_\tau = (\tau-1)p_1+(N-\tau+1)m$. Hence, in this case we obtain $\sigma=n-(N-\tau)(m-p_1)$.
\item $n > (N-1)(m-p_1)$. In this case, from Theorem \ref{less_than_normal_rank} we have that the system matrix is full-column normal rank, namely $n+Nm$, while, according to Proposition \ref{prop:rankDtau}, the rank of $\rank D_\tau$ is still $(\tau-1)p_1+(N-\tau-1)m$. Then we can conclude that $\sigma = (\tau-1)(m-p_1)$.
\end{enumerate}
\end{IEEEproof}

The following corollary studies  zeros at the origin.

\begin{corollary} \label{zerosatzerop1<m}
Consider   the system $\sum_\tau$  $\forall \tau \in \{1,\ldots,N\}$, with  $p_1 \leq m$ and  $Np_1+p_2>Nm$.  Assume that the defining matrices $\{A, B, C^f, D^f, C^s, D^s\}$ take generic values. Then $M_\tau(Z)$    has    zeros at $Z=0$ with multiplicity equal to:
\begin{enumerate}
\item $0$ if $n \leq (\tau-1)(m-p_1)$;
\item $n-(\tau-1)(m-p_1)$ if $(\tau-1)(m-p_1) < n \leq (N-1)(m-p_1)$;
\item $(N-\tau)(m-p_1)$ if $n > (N-1)(m-p_1)$.
\end{enumerate}
\end{corollary}
\begin{IEEEproof}
Pick $\bar \tau$ in the set $\{1,2,\ldots,\,N\}$ and consider the following situations.
\begin{enumerate}
\item $n \leq (N-\bar \tau)(m-p_1)$. In this case, from Theorem \ref{thm:zeroatinfiityp1<m} one can see that the system $\sum_{\bar \tau}$ has no zeros at infinity. Then, recalling Lemma \ref{th:zeros_zero_infinity}, we also have that $\sum_{N-\bar \tau +1}$ has no zeros at $Z=0$. Then, by defining $\tau = N-\bar \tau+1$ and substituting in the inequality $n \leq (N-\bar \tau)(m-p_1)$, one can easily obtain that, when $n \leq (\tau-1)(m-p_1)$, the system  $\sum_{N-\bar \tau +1} \equiv\sum_\tau$ has no zeros at $Z=0$.
\item $(N-\bar \tau)(m-p_1) < n \leq (N-1)(m-p_1)$. In this case, $\sum_{\bar \tau}$ has $n-(N-\bar \tau)(m-p_1)$ zeros at infinity. Using the same arguments employed for the previous case, we can conclude that, when $(\tau-1)(m-p_1) < n \leq (N-1)(m-p_1)$, $\sum_{\tau}$ has $n-(\tau-1)(m-p_1)$ zeros at $Z=0$.
\item $n > (N-1)(m-p_1)$. Again, since $\sum_{\bar \tau}$ has $(\bar \tau-1)(m-p_1)$ zeros at infinity, we have that $\sum_{\tau}$ has  $(N-\tau)(m-p_1)$ zeros at the origin.
\end{enumerate}
\end{IEEEproof}

\begin{remark}
The above results  reveal that, assuming $A,B$, etc. generic  with $p_1 \leq m$ and  $Np_1+p_2>Nm$, when $\tau=1$ all zeros are at the origin and no zero at infinity.  Conversely,  when $\tau=N$ all zeros are at infinity and there are no zeros at the origin. Furthermore, when $\tau=1$ there is always at least one zero at the origin, while when $\tau=N$ there is always at least one zero at infinity (unless one considers a system with no dynamics, i.e. a system with $n=0$).
\end{remark}
\begin{remark}
When $p_1=m$, the conditions given in Theorem \ref{thm:zeroatinfiityp1<m} and the subsequent Corollary on the presence of zeros at $Z=0$ and $Z=\infty$ shrink to empty sets. Then, it follows that $\sum_\tau$ has neither zeros at the origin nor at infinity.
\end{remark}
\begin{remark}
In some special cases depending on the state, input and output  dimensions, $\sum_\tau$ may have zeros at the origin or at infinity for some values of $\tau$ but be completely zero-free for other values of $\tau$. For example,  consider $\sum_\tau$  for particular choice of   $n = 5$, $m = 5$, $p_1 = 3$, $p_2 = 24$ and  $N = 8$ which  has zeros for all values of $\tau$, except for $\tau = 4,\,5$. In these particular cases, the system $\sum_\tau$ is totally zero-free. This can be easily checked by using Theorem \ref{thm:zeroatinfiityp1<m}  and the subsequent Corollary.
\end{remark}

Various theorems have been introduced in this paper regarding zeros of  the system $\Sigma_\tau$ given a generic underlying multirate system. Accordingly, we summarize results obtained in this paper in the  table below.

\begin{table}[h!]
 \caption{Summarizing  the results obtained in this paper.}
\begin{center}
\begin{tabular}{|c|c|c|}
\hline
 \backslashbox{Zero}{Region}&  $p_1 \ge m$  &  {\footnotesize \begin{tabular}[x]{@{}c@{}} $p_1<m$,  \\ $Np_1+p_2>Nm$  \end{tabular} } \tabularnewline
\hline
  Finite nonzero zeros &  No & No \tabularnewline
 \hline
Zeros at zero & No & Zeros can be at these    \tabularnewline
\cline{1-2}
  Zeros at infinity & No  &  \begin{tabular}[x]{@{}c@{}}points  depending on $\tau$. \end{tabular}   \tabularnewline
\hline

\end{tabular}

\label{table:sumtable}
\end{center}
\end{table}

\section{Conclusions}\label{sec:conclusion}
Zeros of tall  discrete-time multirate linear systems were  addressed in this paper, with the zeros  of multirate linear systems being  defined as those of their corresponding blocked systems. The system matrix of tall  blocked systems was investigated for generic choice of parameter matrices. It was specifically shown that  tall  blocked   systems  generically have no finite nonzero zeros.   However,  we showed that there are  situations in which these systems present zeros at $Z=0$ or  $Z=\infty$ or both. Such situations can be characterized in terms of the relevant integer parameters (input, state, and output dimensions and ratio of sampling rates. As part of the investigation, we also identified the generic rank assumed by the system matrix of a blocked system and the transfer function of that system. As  part of  our  future work, we  intend   to generalize  the results of this paper. In particular, we are interested  in a general case  where there are two output streams,  one available  every  $\omega$ time instants   and the other  every $\overline{\omega}$ time instants, with  $\omega$ and $\overline{\omega}$   coprime integers.

\section*{Acknowledgements}
Support by the ARC Discovery Project Grant DP1092571, the FWF (Austrian Science Fund) under contracts P17378 and P20833/N18 and the Oesterreichische Forschungsgemeinschaft is gratefully acknowledged.

\bibliographystyle{plain}
\bibliography{tCONguide}

\begin{thebibliography}{10}

\bibitem{anderson2008}
B.~D.~O. Anderson and M.~Deistler.
\newblock Properties of zero-free transfer function matrices.
\newblock {\em SICE Journal of Control, Measurement and System Integration},
  82(4):284--292, May 2007.

\bibitem{anderonandzamani2012}
B.~D.~O. Anderson, M.~Zamani, and G.~Bottegal.
\newblock On the zero properties of tall linear systems with single-rate and
  multirate outputs.
\newblock In K.~H{\"u}per and J.~Trumpf, editors, {\em Mathematical System
  Theory -- Festschrift in Honor of Uwe Helmke on the Occasion of his Sixtieth
  Birthday}, pages 31--49. CreateSpace, 2013.

\bibitem{bittanti86}
S.~Bittanti.
\newblock Deterministic and stochastic linear periodic systems.
\newblock {\em Lecture Notes in Control and Information Sciences}, 86:141--182,
  1986.

\bibitem{bittanti09}
S.~Bittanti and P.~Colaneri.
\newblock {\em Periodic Systems Filtering and Control}.
\newblock Communications and Control Engineering. Springer-Verlag, 2009.

\bibitem{Bolzern86}
P.~Bolzern, P.~Colaneri, and R.~Scattolini.
\newblock Zeros of discrete-time linear periodic systems.
\newblock {\em IEEE Transactions on Automatic Control}, 31(11):1057 -- 1058,
  November 1986.

\bibitem{chen2010}
W.~Chen, B.~D.~O. Anderson, M.~Deistler, and A.~Filler.
\newblock Properties of blocked linear system.
\newblock {\em Proceedings of the International Federation of Automatic Control
  Conference, 2011. IFAC'11.}, pages 4558--4563, 2011.

\bibitem{Christou2010}
N.~Christou, N.~Karcanias, and M.~Mitrouli.
\newblock The eres method for computing the approximate gcd of several
  polynomials.
\newblock {\em Applied Numerical Mathematics}, 60(1-2):94--114, 2010.

\bibitem{clement2008}
Michael.~P. Clements and Ana.~B Galv\~{a}o.
\newblock Macroeconomic forecasting with mixed-frequency data.
\newblock {\em Journal of Business \& Economic Statistics}, 26:546--554, 2008.

\bibitem{Colaneri95}
P.~Colaneri and S.~Longhi.
\newblock The realization problem for linear periodic systems.
\newblock {\em Automatica}, 31(5):775 -- 779, 1995.

\bibitem{Colaneri1990}
P.~Colaneri, R.~Scattolini, and N.~Schiavoni.
\newblock Stabilization of multirate sampled-data linear systems.
\newblock {\em Automatica}, 26(2):377 -- 380, 1990.

\bibitem{Colaneri92}
P.~Colaneri, R.~Scattolini, and N.~Schiavoni.
\newblock Lqg optimal control of multirate sampled-data systems.
\newblock {\em Automatic Control, IEEE Transactions on}, 37(5):675--682, 1992.

\bibitem{diestler2010}
M.~Deistler, B.~D.~O. Anderson, A.~Filler, Ch. Zinner, and W.~Chen.
\newblock Generalized linear dynamic factor models: An approach via singular
  autoregressions.
\newblock {\em European Journal of Control}, 3:211--224, 2010.

\bibitem{filler2010}
A.~Filler.
\newblock {\em Generalized Dynamic Factor Models Structure Theory and
  Estimation for Single Frequency and Mixed Frequency Data}.
\newblock PhD thesis, Vienna University of Technology, 2010.

\bibitem{fillerthesis}
A.~Filler.
\newblock {\em Generalized dynamic factor models structure theory and
  estimation for single frequency and mixed frequency data}.
\newblock PhD thesis, Vienna University of Technology, 2010.

\bibitem{Forni2000}
M.~Forni, M.~Hallin, M.~Lippi, and L.~Reichlin.
\newblock The generalized dynamic-factor model: Identification and estimation.
\newblock {\em The Review of Economics and Statistics}, 82(4):540--554,
  November 2000.

\bibitem{Grasselli88}
O.~M. Grasselli and S.~Longhi.
\newblock Zeros and poles of linear periodic multivariable discrete-time
  systems.
\newblock {\em Circuits, Systems, and Signal Processing}, 7:361--380, 1988.

\bibitem{Ltkepohlbook}
L.~Helmut.
\newblock {\em New Introduction to Multiple Time Series Analysis}.
\newblock Springer, 2007.

\bibitem{hespanha_book}
J.~P. Hespanha.
\newblock {\em Linear Systems Theory}.
\newblock Princeton University Press, 2009.

\bibitem{kailath}
T.~Kailath.
\newblock {\em Linear Systems}.
\newblock Prentice-Hall, New Jersey, 1980.

\bibitem{karsanias1979}
N.~Karcanias and B.~Kouvaritakis.
\newblock The output zeroing problem and its relationship to the invariant zero
  structure : a matrix pencil approach.
\newblock {\em International Journal of Control}, 30(3):395--415, 1979.

\bibitem{karsanias2002}
N.~Karcanias and D.~Vafiadis.
\newblock Canonical forms for state-space descriptions.
\newblock {\em Control Systems, Robotics and Automation}, 5:361--380, 2002.

\bibitem{Khargoneckar85}
P.~Khargonekar, K.~Poolla, and A.~Tannenbaum.
\newblock Robust control of linear time-invariant plants using periodic
  compensation.
\newblock {\em IEEE Transactions on Automatic Control}, 30(11):1088 -- 1096,
  November 1985.

\bibitem{Mitrouli93}
M.~Mitrouli and N.~Karcanias.
\newblock Computation of the gcd of polynomials using gaussian transformations
  and shifting.
\newblock {\em International Journal of Control}, 58(1):211--228, 1993.

\bibitem{Raknerud2007}
A.~Raknerud, T.~Skjerpen, and A.~R. Swensen.
\newblock Forecasting key macroeconomic variables from a large number of
  predictors: A state space approach.
\newblock Discussion Papers 504, Research Department of Statistics Norway, May
  2007.

\bibitem{rosenbrock1974}
H.~H. Rosenbrock.
\newblock {\em Computer-Aided Design of Control Systems}.
\newblock Cambridge, London, 1974.

\bibitem{Schumacher2006}
C.~Schumacher and J.~Breitung.
\newblock Real-time forecasting of {GDP} based on a large factor model with
  monthly and quarterly data.
\newblock Discussion Paper Series 1: Economic Studies 2006,33, Deutsche
  Bundesbank, Research Centre, 2006.

\bibitem{chenB95}
C.~Tongwen and B.~A. Francis.
\newblock {\em Optimal Sampled-Data Control Systems}.
\newblock Springer-Verlag New York, Inc., Secaucus, NJ, USA, 1995.

\bibitem{vaid93}
P.~P. Vaidyanathan.
\newblock {\em {Multirate Systems and Filter Banks}}.
\newblock Prentice-Hall, Inc., Upper Saddle River, NJ, USA, 1993.

\bibitem{vandooren1979}
P.~{Van Dooren}.
\newblock The computation of {Kronecker's} canonical form of a singular pencil.
\newblock {\em Linear Algebra and Its Applications}, 27:103--140, 1979.

\bibitem{wonhem1979}
W.~M. Wonham.
\newblock {\em Linear multivariable control: a geometric approach}.
\newblock Springer-Verlag, New York, 1979.

\bibitem{zamani2012acc}
M.~Zamani and B.~D.~O. Anderson.
\newblock One the zero properties of linear discrete-time systems with
  multirate outputs.
\newblock {\em Proceedings of the 2012 American Control Conference}, pages
  5182--5187, 2012.

\bibitem{zamaniscl}
M.~Zamani, B.~D.~O. Anderson, U.~Helmke, and W.~Chen.
\newblock On the zeros of blocked time-invariant systems.
\newblock {\em Systems \& Control Letters}, 62(7):597 -- 603, 2013.

\bibitem{zamani2011}
M.~Zamani, W.~Chen, B.~D.~O. Anderson, M.~Deistler, and A.~Filler.
\newblock On the zeros of blocked linear systems with single and mixed
  frequency data.
\newblock {\em Proceedings of the 2011 Control and Desicion Conference}, pages
  4312--4317, 2011.

\bibitem{Zhou1996}
K.~Zhou, J.~C. Doyle, and K.~Glover.
\newblock {\em Robust and Optimal Control}.
\newblock Prentice-Hall, Inc., Upper Saddle River, NJ, USA, 1996.

\end{thebibliography}

\appendix
\subsection{Proof of Proposition \ref{prop:rankDtau}}
We first need to introduce the following lemma.
\begin{lemma} \label{generic_controllability}
Consider a generic pair of matrices $A \in \mathbb{R}^{n \times n}$ and $B \in \mathbb{R}^{n \times m}$. Then, given $\nu \in \mathbb{N}$, the matrix
\begin{equation} \label{contr_matr}
\mathcal{C} = \begin{bmatrix} B & AB & \ldots & A^{\nu-1}B \end{bmatrix}
\end{equation}
is always full rank, i.e. its rank is equal to:
\begin{enumerate}
\item its number of rows, $n$, if $n \leq \nu m$,
\item its number of columns, $\nu m$, if $n > \nu m$.
\end{enumerate}
\end{lemma}
\begin{IEEEproof}
Since the case $m \geq n$ is straightforward, we focus on the case $n>m$. The statement can be proven by finding a pair $(A,\,B)$ such that the matrix $\mathcal{C}$ attains full rank, since it means that this happens for any generic pair of such matrices. Accordingly, choose the following matrices
\begin{equation}
A = \begin{bmatrix}
 0_{m \times (n-m)} & I_{m} \\
I_{n-m} & 0_{(n-m) \times m} \end{bmatrix} \qquad
B = \begin{bmatrix}
I_{m}  \\
0_{(n-m)\times m} \end{bmatrix} \,,
\end{equation}
regarding which we point out the following properties.
\begin{enumerate}
\item The matrix $A$ acts as a circular left-shift operator matrix through $m$ positions and can be written in terms of the canonical basis of $\mathbb{R}^n$, as $A = \begin{bmatrix} e_{m+1} & \ldots & e_{n} & e_1 & \ldots & e_{m} \end{bmatrix}$. Then, if for example $n > 3m+1$, one has $A^2 = \begin{bmatrix} e_{2m+1} & \ldots & e_{n}  & e_1 &  \ldots & e_{2m}  \end{bmatrix}$, $A^3 = \begin{bmatrix} e_{3m+1} & \ldots & e_{n}  & e_1 &  \ldots & e_{3m}  \end{bmatrix}$.
\item The matrix $B$ selects the first $m$ columns of any matrix which premultiplies it. Furthermore, the columns of $B$ correspond to $e_1,\,\ldots,\,e_{m}$.
\end{enumerate}
Based on these considerations, we have then
\begin{eqnarray*}
B  &= & \begin{bmatrix} e_1 & \ldots & e_{m} \end{bmatrix} \\
AB &= &\begin{bmatrix} e_{m+1} & \ldots & e_{2m} \end{bmatrix} \\
A^2B &  = &\begin{bmatrix} e_{2m+1} & \ldots & e_{3m}  \end{bmatrix} \\
&\vdots&\\
A^{\nu-1}B & = &\begin{bmatrix} e_{(\nu-1)m+1} & \ldots & e_{\nu m} \end{bmatrix}\,,
\end{eqnarray*}
where for simplicity we have adopted the notation $e_{(kn+i)} = e_i$, $i = 1,\,\ldots,\,n$, $k \in \mathbb{N}$. Then, it is easy to conclude that:
\begin{enumerate}
\item if $n \leq \nu m$, all the vectors of the canonical basis of $\mathbb{R}^n$ enter in the matrix $\mathcal{C}$ at least once, and thus $\mathcal{C}$ is full row rank;
\item if $n > \nu m$, there are $\nu m$ distinct vectors of the canonical basis of $\mathbb{R}^n$ entering in the matrix $\mathcal{C}$ and thus $\mathcal{C}$ is full column rank.
\end{enumerate}
\end{IEEEproof}

We can now prove Proposition \ref{prop:rankDtau}.
For the sake of brevity, we treat only the case $n \geq m$, since the case $n < m$ is virtually the same. Fix $\tau$ and first assume $n \leq (N-\tau)(m-p_1)$. We  consider a particular system, defined by the matrices
\begin{eqnarray} \label{eq:particular_sys}
&A = \begin{bmatrix}
0_{(m-p_1) \times (n-m+p_1)} & I_{m-p_1} \nonumber\\
I_{n-m+p_1} & 0_{(n-m+p_1) \times (m-p_1)} \end{bmatrix} \\
&B = \begin{bmatrix}
\begin{matrix} I_{m-p_1} & 0_{(m-p_1) \times p_1} \end{matrix} \\
0_{(n-m+p_1)\times m} \end{bmatrix} \nonumber\\
&C^f = 0_{p_1 \times n} \quad D^f = \begin{bmatrix} 0_{p_1 \times (m-p_1)} & I_{p_1}  \end{bmatrix}  \\
&C^s = \begin{bmatrix} I_n \\ 0_{(p_2-n) \times n} \end{bmatrix} \quad D^s = \begin{bmatrix} 0_{n \times m} \\ \begin{matrix}  I_{m-p_1} & 0_{(m-p_1) \times  p_1}  \end{matrix} \\ 0_{(p_2+p_1-n-m) \times m} \end{bmatrix}  \nonumber
\end{eqnarray}
Note that, under the working assumptions, the dimensions of the various matrices involved in the construction of such system are consistent. In particular, since $n \leq (N-\tau)(m-p_1)$ and, by assumption of tallness, $p_2 > N(m-p_1)$, one has
\begin{align*}
n+m &\leq (N-\tau)(m-p_1)+m \leq (N-1)(m-p_1)+m \\
    & \leq p_1+N(m-p_1) < p_1 + p_2
\end{align*}
and so $p_2+p_1-n-m > 0$. Below, we adopt the notation that, if a submatrix has zero rows or columns, then it does not appear in the relative matrix.
Before writing $D_\tau$ explicitly, we focus on the submatrix
$$
\begin{bmatrix} C^sA^{N-\tau-1}B & C^sA^{N-\tau-2}B & \ldots & C^sB  \end{bmatrix} \,,
$$
which enters in the block row associated with the slow dynamics of the blocked system.
Due to the structure of $C^s$, a first rewriting yields
\begin{equation} \label{eq:AB}
\begin{bmatrix} A^{N-\tau-1}B & A^{N-\tau-2}B & \ldots & B \end{bmatrix}\,.
\end{equation}
Now, we point out the following properties of $A$ and $B$.
\begin{enumerate}
\item The matrix $A$ acts as a circular left-shift operator matrix through $m-p_1$ positions. 
    Furthermore, the columns of $A$ are orthogonal.
\item The matrix $B$ selects the first $m-p_1$ columns of any matrix which premultiplies it. The other $p_1$ columns of the resulting matrix are set to zero. Furthermore, the nonzero columns of $B$ correspond to $e_1,\,\ldots,\,e_{m-p_1}$.
\end{enumerate}
Based on these considerations, we have then
\begin{eqnarray*}
B \!\! &\!\!=\!\! &\!\!\! \begin{bmatrix} e_1 & \ldots & e_{m-p_1} & 0_{n \times p_1} \end{bmatrix} \\
AB \!\!&\!\! =\!\! &\!\!\! \begin{bmatrix} e_{m-p_1+1} & \ldots & e_{2(m-p_1)} & 0_{n \times p_1} \end{bmatrix} \\
A^2B \!\!&\!\! =\!\! &\!\!\! \begin{bmatrix} e_{2(m-p_1)+1} & \ldots & e_{3(m-p_1)} & 0_{n \times p_1} \end{bmatrix} \\
 &\vdots & \\
A^{N-\tau-1}B \!\!&\!\! =\!\!& \!\!\!\begin{bmatrix} e_{(N-\tau-1)(m-p_1)+1} &\!\!\! \ldots\!\!\! & e_{(N-\tau)(m-p_1)} &\!\!\! 0_{n \times p_1} \end{bmatrix}
\end{eqnarray*}
where for simplicity we have adopted the notation $e_{(kn+i)} = e_i$, $i = 1,\,\ldots,\,n$, $k \in \mathbb{N}$. Thus, since we assumed $n \leq (N-\tau)(m-p_1)$,  the above matrix has rank equal to $n$. Defining $E_i := \begin{bmatrix} e_{(i-1)(m-p_1)+1} & \ldots & e_{i(m-p_1)}\end{bmatrix}$, we can write $D_\tau  =  $
\begin{equation*}
 \left[\begin{array}{cccccccccccc}
\! \! \!               0_{p_1 \times (m-p_1)} \! \! \! &\! \! \!\! \! \!    I_{p_1}                  &\! \! \!                             \! \! \!       &\! \! \!                      &                                                               \\
\! \! \!                                      \! \! \! &\! \! \!\! \! \!       \ddots                &\! \! \!                             \! \! \!       &\! \! \!                      &                                                               \\
\! \! \!                                     \! \! \!  &\! \! \!\! \! \!                             &\! \! \!       0_{p_1 \times (m-p_1)}\! \! \!       &\! \! \!    I_{p_1}           &                                                               \\
\! \! \!                                     \! \! \!  &\! \! \!\! \! \!                             &\! \! \!                             \! \! \!       &\! \! \!                      & \begin{matrix} 0_{p_1 \times (m-p_1)} & I_{p_1} \end{matrix}  \\
\! \! \!                                     \! \! \!  &\! \! \!\! \! \!                             &\! \! \!                             \! \! \!       &\! \! \!                      &                                                               \\
\! \! \!                                     \! \! \!  &\! \! \!\! \! \!                             &\! \! \!                             \! \! \!       &\! \! \!                      &                                                               \\
\! \! \!                                     \! \! \!  &\! \! \!\! \! \!                             &\! \! \!                             \! \! \!       &\! \! \!                      &                                                               \\
\! \! \!                 E_{N-\tau}          \! \! \!  &\! \! \!\! \! \!    0_{n\times p_1}          &\! \! \!    \ldots   E_{1}           \! \! \!       &\! \! \!    0_{n\times p_1}   & 0_{n \times m}  \!  \! \!                                     \\
\! \! \!                                     \! \! \!  &\! \! \!\! \! \!                             &\! \! \!                             \! \! \!       &\! \! \!                      & \begin{matrix} I_{m-p_1} & 0_{(m-p_1) \times p_1} \end{matrix}\\
\! \! \!                                     \! \! \!  &\! \! \!\! \! \!                             &\! \! \!                             \! \! \!       &\! \! \!                      & 0_{(p_2+p_1-n-m) \times m}  \! \! \!                           \end{array} \right. \!\!\!
\end{equation*}
\begin{equation}
\cdots\qquad \left.\begin{array}{cccccccccccc}
\! \! \!\! \! \!                              &           & \! \! \!                           &    \! \! \!   \\
\! \! \!\! \! \!                              &           & \! \! \!                           &    \! \! \!   \\
\! \! \!\! \! \!                              &           & \! \! \!                           &    \! \! \!   \\
\! \! \!\! \! \!                              &           & \! \! \!                           &    \! \! \!   \\
\! \! \!\! \! \!     0_{p_1 \times (m-p_1)}   & I_{p_1}   &  \! \! \!                          &     \! \! \!   \\
\! \! \!\! \! \!                              & \ddots    &  \! \! \!                          &          \! \! \!   \\
\! \! \!\! \! \!                              &           &  \! \! \! 0_{p_1 \times (m-p_1)}   & I_{p_1} \! \! \!   \\
\! \! \!\! \! \!                              &           &  \! \! \!                          &          \! \! \!   \\
\! \! \!\! \! \!                              &           &  \! \! \!                          &          \! \! \!   \\
\! \! \!\! \! \!                              &           &\! \! \!                            &     \! \! \!    \end{array} \right] \,.
\end{equation}

This expression reveals that the rank of $D_\tau$ can be calculated by summing the ranks of each nonzero submatrix entering it. More precisely, we have $N$ identity matrices of size $p_1$ and one identity matrix of size $m-p_1$, plus the $E_i$'s which provide $n$ linearly independent columns in total. Hence, for this choice of parameter matrices and $n \leq (N-\tau)(m-p_1)$ we have $\rank D_\tau = (N-1)p_1 + m + n$. We conclude that, for generic choice of parameter matrices, under these assumptions, $\rank D_\tau \geq (N-1)p_1 + m + n$. 


Now, still assuming  $n \leq (N-\tau)(m-p_1)$  we seek an upper bound for the generic rank of $D_\tau$ and show that indeed it coincides with the lower bound just found. For this, assume generic parameter matrices and introduce the matrix $\bar D_{\tau} \triangleq$
\begin{align}
&\left[\!\!\!\begin{array}{ccc|ccc} D^f & 0  & \ldots & 0 & \ldots & 0\\
\vdots & \ddots &   & \vdots & & \vdots \\
C^fA^{N-\tau-1}B & \ldots  &D^f & 0 & \ldots & 0 \\
C^sA^{N-\tau-1}B & \ldots  &D^s & 0 & \ldots & 0\\
\hline
C^fA^{N-\tau}B  & \ldots & C^fB & D^f & 0 & 0 \\
 \vdots  &  & & & \ddots& \\
C^fA^{N-2}B & \ldots  & C^fA^{\tau-2}B & C^fA^{\tau-3}B & \ldots& D^f \end{array}\right] \nonumber\\
& \triangleq \left[\begin{array}{c|c} \Delta_1 & 0 \\ \hline * & \Delta_2 \end{array}\!\!\!\right]
\end{align}
which, being just a row permutation of $ D_{\tau}$, has the same rank. Hence, from now on we shall refer to the rank of $ D_{\tau}$. The presence of the fat matrix $D^f$ on the block diagonal of $\Delta_2 \in \mathbb{R}^{(\tau -1)p_1 \times (\tau -1)m}$ ensures that $\Delta_2$ is full row rank, namely $\rank(\Delta_2) = (\tau -1)p_1$. This implies that the matrix indicated as ``$*$'' does not influence the rank of $D_\tau$. Thus $\rank(D_\tau) = \rank(\Delta_1) + \rank(\Delta_2)$ and so we focus on $\Delta_1$. We define
\begin{align*}
\Delta_a &\triangleq \left[\begin{array}{cccc} D^f & 0 & \ldots & 0\\
C^fB & D^f & \ldots & 0 \\
\vdots & \vdots & \ddots & \vdots \\
C^f A^{N-\tau-2}B & \ldots & D^f & 0 \end{array}\right] \\
\Delta_b &\triangleq \left[\begin{array}{cccc}
C^fA^{N-\tau-1}B & \ldots & C^fB &D^f  \\
C^sA^{N-\tau-1}B & \ldots & C^sB &D^s  \end{array}\right]
\end{align*}
so that
$$
\Delta_1 = \begin{bmatrix} \Delta_a    \\ \Delta_b  \end{bmatrix} \,
$$
and $\rank(\Delta_1) \leq \rank(\Delta_a) + \rank(\Delta_b)$. Note that, $\Delta_1$ is a tall matrix, since it includes the slow rate outputs whose dimension ensure tallness in the whole system. Hence, its maximum achievable rank is given by the number of its columns, namely $(N-\tau+1)m$. Thus we can find a first upper bound for the rank of $D_\tau$, that is
\begin{equation} \label{max_rank_D}
\rank(D_\tau) \leq (N-\tau+1)m + (\tau-1)p_1 \,,
\end{equation}
and this will be used below. Meanwhile, we focus on the analysis of $\Delta_a$. It is well-known (see e.g. \cite{zamani2011}) that, due to genericity of the matrix $D^f$, $\Delta_a$ is full row rank, namely $(N-\tau)p_1$. For  $\Delta_b$, we consider the following factorization
\begin{align} \label{eq:Delta_b}
\Delta_b & = \begin{bmatrix}
C^f & D^f \\
C^s & D^s
\end{bmatrix} \begin{bmatrix} A^{N-\tau-1}B & A^{N-\tau-2}B & \ldots & B & 0 \\
                                0 & 0 & \ldots & 0 & I_m \end{bmatrix} \nonumber\\ 
 & \triangleq H \mathcal{R} \,.
\end{align}
Since by assumption $n \leq (N-\tau)(m-p_1)$, from Lemma \ref{generic_controllability} one can see that the matrix $\mathcal{R}$ is full row rank, namely $n+m$. Thus, the rank of $\Delta_b$ is determined by $H \in \mathbb{R}^{(p_1+p_2) \times (n+m)}$. On the one hand, assumption of tallness of the blocked system ensures $p_2 > N(m-p_1)$; on the other hand, since $n \leq (N-\tau)(m-p_1)$, one has $n+m < p_2 + p_1$. Hence $\Delta_b$ is tall, and so  generically $\rank(\Delta_b) = n+m$ and $\rank(\Delta_1) \leq \rank(\Delta_a) + \rank(\Delta_b) = (N-\tau)p_1 + n +m$, which in turn implies 
\begin{align*}
\rank(D_\tau) & = \rank(\Delta_1) + \rank(\Delta_2) \\
        &   \leq (N-\tau)p_1 + n +m + (\tau -1)p_1 \\
        &   = (N-1)p_1 + n +m \,,
\end{align*}
which corresponds to the lower bound found previously.

In order to complete our proof, it remains to analyze the case $n > (N-\tau)(m-p_1)$. To do so, we first make an observation concerning the case $n = (N-\tau)(m-p_1)$, which was covered in the first part of the proof. In this particular case, $\rank(D_\tau) = (N-1)p_1 + n +m = (N-\tau+1)m + (\tau-1)p_1$, which corresponds to the upper bound on rank of $D_\tau$ given by \eqref{max_rank_D}. Now, the proof for the case $n > (N-\tau)(m-p_1)$ can be completed by showing that such an upper bound is attained by any generic tall system with $n = (N-\tau)(m-p_1) + q$, $q \in \mathbb{N}$. This can be verified by choosing the system
\begin{eqnarray} \label{eq:particular_sys_2}
&A = \left[\begin{matrix}
0_{(m-p_1) \times (N-\tau-1)(m-p_1)} &  I_{m-p_1} \\
I_{(N-\tau-1)(m-p_1)} & 0_{(N-\tau-1)(m-p_1) \times (m-p_1)}\\ 
0_{q \times ((N-\tau-1)(m-p_1)}  &0_{q \times (m-p_1)}   \end{matrix} \right. \nonumber\\
&\cdots\,\,\left. \begin{matrix}
 0_{(m-p_1) \times q}\\
 0_{((N-\tau-1)(m-p_1) \times q} \\
0_{q\times q}  \end{matrix} \right] \nonumber\\ 
&B = \begin{bmatrix}
\begin{matrix} I_{m-p_1} & 0_{(m-p_1) \times p_1} \end{matrix} \\
0_{(n-m+p_1)\times m} \end{bmatrix}
\end{eqnarray}
\begin{eqnarray*}
&C^f = 0_{p_1 \times n} \quad D^f = \begin{bmatrix} 0_{p_1 \times (m-p_1)} & I_{p_1}  \end{bmatrix} \nonumber \\
&C^s = \begin{bmatrix} \begin{matrix} I_{(N-\tau)(m-p_1)} & 0_{(N-\tau)(m-p_1) \times q } \end{matrix}  \\ 0_{(p_2-(N-\tau)(m-p_1)) \times n} \end{bmatrix} \nonumber \\
&D^s = \begin{bmatrix} 0_{(N-\tau)(m-p_1) \times m} \\ \begin{matrix}  I_{m-p_1} & 0_{(m-p_1) \times  p_1} \end{matrix} \\ 0_{(p_2+p_1-m-(N-\tau)(m-p_1)) \times m} \end{bmatrix} \,, \nonumber
\end{eqnarray*}
which generates a matrix $D_\tau$ equal to the one generated by the system \eqref{eq:particular_sys}, when $n = (N-\tau)(m-p_1)$. Since we have previously proven that, in that case, the rank is $(\tau-1)p_1 + (N-\tau+1)m$ (which is also the maximum rank achievable), then also for any $n > (N-\tau)(m-p_1)$ we have $\rank(D_\tau) = (\tau-1)p_1 + (N-\tau+1)m$. This completes the proof.

\subsection{Proof of Theorem \ref{less_than_normal_rank}}
Before proving our result on the normal rank of  $M_\tau(Z)$, we need to introduce three preliminar results. The following lemma is adopted from \cite{bittanti09} and modified for our own  purpose.
\begin{lemma} \label{relation between different transfer functions}
The transfer function $V_ \tau(Z)$ associated  with the blocked system  (\ref{blockedsystem1}) has the following property
\begin{equation}\label{eq:tauandtau+1}
\begin{split}
V_{\tau+1}(Z)=&\\
&\!\!\!\!\!\!\!\!\!\!\!\!\!\!\!\!\!\!\!\!\!\!\!\left[\begin{array}{ccc}
0 & I_{p_{1(N-1)}} & 0\\
ZI_{p_{1}} & 0 & 0\\
0 & 0 & I_{p_{2}}\end{array}\right]V_{\tau}(Z)\left[\begin{array}{cc}
0 & Z^{-1}I_{m}\\
I_{m(N-1)} & 0\end{array}\right],
\end{split}
\end{equation}
where $\tau\in\{1,2\ldots,N-1\}$.
\end{lemma}
\begin{proposition} \label{th:constant_normal_rank}
The normal rank of the system matrix $M_\tau(Z)$ is same  for every value of $\tau \in  \{1, 2,\ldots ,N\}$.
\end{proposition}
\begin{IEEEproof}
Using the above lemma, one can easily conclude that  the transfer function matrices  $V_{\tau+1}(Z_0)$ and  $V_\tau(Z_0)$  have the same rank provided that  $Z_0$ does not belong to the finite set of poles of the $V_\tau(Z)$ (which is  the same as that of $V_{\tau+1}(Z)$) and $Z_0 \notin \{0,\infty\}$. Hence, we can conclude that  $V_{\tau+1}(Z)$ and  $V_\tau(Z)$ have the same normal rank and so do their associated system matrices i.e. $M_{\tau+1}(Z)$ and $M_{\tau}(Z)$.

\end{IEEEproof}

\begin{proposition} \label{prop:rankD1}
Consider the system $\sum_1$ (i.e. the blocked system obtained with $\tau = 1$), with
$p_1 < m$, $Np_1+p_2>Nm$ and generic values of the defining matrices
$\{A,\,B,\,C^f,\,C^s,\,D^f,\,D^s\}$. Then:
\begin{enumerate}
\item if $n \leq (N-1)(m-p_1)$, the matrix $D_1$ has rank equal to $(N-1)p_1+m+n$;
\item if $n > (N-1)(m-p_1)$, the matrix $D_1$ has full-column rank, namely $Nm$.
\end{enumerate}
\end{proposition}
\begin{IEEEproof}
The proof follows easily from Proposition \ref{prop:rankDtau}, by letting $\tau = 1$.
\end{IEEEproof}

We are now ready to prove Theorem \ref{less_than_normal_rank}. Here, we focus on the matrix
$M_1(Z)$; every result on its normal rank can be easily extended to
any value of $\tau = \{2,\,\ldots,\,N\}$ using Proposition
\ref{th:constant_normal_rank}.

Consider the matrix $D_1$ and define $r \triangleq \rank (D_1)$; note that the condition of tallness of the system implies $r \leq Nm$.
Define the full row rank matrix $\bar D_1 \in \mathbb{R}^{r\times Nm}$, obtained by discarding a proper number of linearly dependent rows of
$D_1$. Similarly, define $\bar C_1$ discarding the corresponding
rows from $C_1$. Without loss of generality assume $A$ diagonal.
This hypothesis is not limiting; in fact, under a generic setting,
$A$ has $n$ distinct eigenvalues and so it is diagonalizable. If one
considers a change of basis $T$ such that $T^{-1}A T$ is diagonal,
then the other parameter matrices $T^{-1}B$ and $C T$ are still in
a generic setting. Define $\bar M_1(Z)$ as follows
\begin{equation}
\bar M_1(Z) = \begin{bmatrix} Z - a_1^N  & \ldots & 0 & -b^T_1 \\
                            \vdots &\ddots&\vdots &\vdots\\
                            0 &  \ldots  & Z - a_n^N & -b^T_n \\
                            \bar c_{1,1}   & \ldots & \bar c_{1,n} & \bar D_1 \end{bmatrix} \,,
\end{equation}
where the $a_i$'s represent the diagonal elements of $A$, $b^T_i$ is
the i-th row of $B_1$ and $ \bar c_{i,1}$ is the i-th column of
$\bar C_1$. Consider the submatrix $\begin{bmatrix} \bar c_{1,n} &
\bar D_1 \end{bmatrix}$. Since $\bar D_1$ is full row rank, also
this matrix is full row rank. Consider the equation
\begin{equation}
v^T \begin{bmatrix} \bar c_{1,n} & \bar D_1 \end{bmatrix} =
\begin{bmatrix}  Z - a_n^N & -b^T_n \end{bmatrix} \,,
\end{equation}
in which $v$ and $Z$ are yet to be specified and which can be rewritten as
\begin{equation}
\left\{ \begin{array}{lcl}
v^T \bar c_{1,n} & = &  Z - a_n^N  \\
v^T \bar D_1 & = & -b^T_n \\
\end{array} \right. \,.
\end{equation}
Since $\bar D_1$ is full row rank there exists at most one vector
$\bar v^T$ satisfying the second relation. Clearly, if one were to insert such a
vector in the first relation,  there could exist only one value $Z_n \in
\mathbb{C}$ such that this equation is satisfied. Choose $Z \neq
Z_n$ and consider the submatrix
\begin{equation}
\begin{bmatrix}
  0 & Z - a_n^N & -b^T_n \\
\bar c_{1,n-1} & \bar c_{1,n} & \bar D_1
\end{bmatrix} \,,
\end{equation}
which is clearly full row rank, namely $r + 1$.
Write the equation
\begin{equation}
v^T \begin{bmatrix}
  0 & Z - a_n^N & -b^T_n \\
\bar c_{1,n-1} & \bar c_{1,n} & \bar D_1
\end{bmatrix} =  \begin{bmatrix}  Z - a_{n-1}^N & 0 & -b^T_{n-1} \end{bmatrix} \,,
\end{equation}
which in turn can be rewritten as
\begin{equation}
\left\{ \begin{array}{lcl}
v^T \begin{bmatrix} 0 \\ \bar c_{1,n-1} \end{bmatrix} & = &  Z - a_{n-1}^N \\
&&\vspace{-0.2cm}\\
v^T \begin{bmatrix} Z - a_n^N & -b^T_n \\ \bar c_{1,n} & \bar D_1
\end{bmatrix} & = &  \begin{bmatrix} 0 & -b^T_{n-1} \end{bmatrix}
\end{array} \right. \,.
\end{equation}
Again, the second relation admits at most one solution, which is
compatible with the first equation for only one value $Z_{n-1} \in
\mathbb{C}$. Hence, choosing $Z \notin \{Z_n,\,Z_{n-1}\}$ one can
build the matrix
\begin{equation}
\begin{bmatrix}
 0 &  Z - a_{n-1}^N &  0  & -b^T_{n-1} \\
 0 &  0 & Z - a_n^N & -b^T_n \\
\bar c_{1,n-2} & \bar c_{1,n-1} & \bar c_{1,n} & \bar D_1
\end{bmatrix} \,,
\end{equation}
which is full row rank, namely $r + 2$, and repeat
the previous steps until all the rows containing the $a^N_i$'s and
the $b_i^T$'s, $i \in \{1,\ldots,\,n\}$, are considered. This
procedure ends after $n$ iterations, when all the rows of the matrix
$\bar M_1(Z)$ are included; clearly the rank turns out to be $r+n$. Since $\bar M_1(Z)$ is a submatrix of $M_1(Z)$, the
normal rank of $M_1(Z)$ is greater than or equal to $r+n$.
There are two cases in the theorem statement. Treating the second one first, suppose
$n \geq (N-1)(m-p_1)$. Recalling Proposition \ref{prop:rankD1}, $r = Nm$; hence $\nrank (\bar M_1(Z)) = n+Nm$ and $M_1(Z)$ is full normal rank.

For the second case, suppose $n < (N-1)(m-p_1)$. In this case, from Proposition \ref{prop:rankD1} we have $r = (N-1)p_1+m+n$, hence $\nrank  (M_1(Z)) \geq \nrank (\bar M_1(Z)) = (N-1)p_1+m+2n$.
Now, consider the submatrix formed by the first $n+(N-1)p_1$ rows of
$M_1(Z)$. Such a submatrix is full normal rank, since it can be seen
also as a submatrix of the system matrix
\begin{equation}\label{eq:fat1}
\left[\begin{array}{cccc}
ZI_n-A^N&-A^{N-1}B&\dots&-B\\
C^f&D^f&&0\\
\vdots&\vdots&\ddots&\vdots\\
C^fA^{N-1}&C^fA^{N-2}B&\dots&D^f
\end{array}\right]\,,
\end{equation}
which is the system matrix of a blocked fat system with generic
parameter matrices. From \cite{zamani2011}, it is
well-known that \eqref{eq:fat1} is full normal rank. Now consider
the remaining rows of $M_1(Z)$, i.e. the matrix
\[
\Pi=\left[\begin{array}{ccccc}
C^{f}A^{N-1} & C^{f}A^{N-2}B & \ldots & C^{f}B & D^{f}\\
C^{s}A^{N-1} & C^{s}A^{N-2}B & \ldots & C^{s}B &
D^{s}\end{array}\right]
\]
which can be factorized as
\[
\Pi=\left[\begin{array}{cc}
C^{f}  &  D^{f}\\
C^{s} & D^{s}\end{array}\right] \begin{bmatrix} A^{N-1} & A^{N-2}B & \ldots & B & 0 \\
                                0 & 0 & \ldots & 0 & I_m \end{bmatrix} \triangleq H \bar{\mathcal{R}} \,.
\]
Since $A$ is full rank, then also $A^{N-1}$ is full rank and thus the matrix $\bar{\mathcal{R}}$ is full row rank, namely $n+m$. Thus, the rank of $\Pi$ depends on the rank of $H$, which, for generic choice of  matrices $C^s,D^s,C^f,D^f$,  is equal to $\alpha \triangleq \min\{p_1+p_2, m+n\}$. Then $\nrank (M_1(Z)) \leq n+(N-1)p_1+\alpha$. However, since for
the condition of tallness $p_2 > N(m-p_1)$ and by assumption
$n<(N-1)(m-p_1)$, we have
$
n+m  < (N-1)(m-p_1) + m = N(m - p_1) + p_1 < p_2 + p_1 \,,
$
and so $ \alpha = n+m$. Hence
$\nrank (M_1(Z)) \leq (N-1)p_1+m+2n$. Combining this bound
with the lower bound found previously, we conclude that
$\nrank (M_1(Z)) = (N-1)p_1+m+2n$.

\end{document}